\documentclass{aa}
\usepackage{graphicx}
\usepackage{natbib,twoopt}
\usepackage{txfonts}
\usepackage{ textcomp }
\bibpunct{(}{)}{;}{a}{}{,} % to follow the A&A style
\newcommand{\rcra}{R~CrA}

\usepackage{hyperref}
   \title{The origin of R~CrA variability }

   \subtitle{A complex triple system hosting a disk}

   \author{E. Sissa\inst{1},
           R. Gratton\inst{1},
           J.M. Alcal\`a\inst{2},
           S. Desidera\inst{1},
           S. Messina\inst{3},
           D. Mesa\inst{1},
           V. D'Orazi\inst{1},
           E. Rigliaco\inst{1}
          }
\authorrunning{E. Sissa et al.}
\offprints{E. Sissa,  \\
   \email{elena.sissa@inaf.it} }
   
\institute{INAF-Osservatorio Astronomico di Padova,  Vicolo dell'Osservatorio 5, I-35122, Padova, Italy \and
INAF-Osservatorio Astronomico di Capodimonte, Salita Moiariello 16, I-80131, Napoli, Italy   \and
INAF-Osservatorio Astrofisico di Catania, Via S.Sofia 78, I-95123, Catania, Italy}

\date{Received  /
Accepted}

  \abstract
   {R~CrA is the brightest member of the Coronet star forming region and it is the closest Herbig AeBe star with a spectrum dominated by emission lines. Its luminosity has been monitored since the end of the 19th century, but the origin of its variability, which shows a stable period of $65.767\pm 0.007$~days, is still unknown.  }
   {We studied photometric and spectroscopic data for this star to investigate the nature of the variability of R~CrA.}
   {We exploited the fact that near infrared luminosity of the Herbig AeBe stars is roughly proportional to the total luminosity of the stars to derive the absorption, and then mass and age of R~CrA. In addition, we model the periodic modulation of the light curve as due to partial attenuation of a central binary by a circumbinary disk. This model reproduces very well the observations.}
   {We found that the central object in R~CrA is a very young ($1.5\pm 1.5$~Myr), highly absorbed ($A_V=5.47\pm 0.4$~mag) binary; we obtain masses of $M_A=3.02\pm 0.43$~M$_\odot$ and $M_B=2.32\pm 0.35$~M$_\odot$ for the two components. We propose that the secular decrease of the R~CrA apparent luminosity is due to a progressive increase of the disk absorption. This might be related to precession of a slightly inclined disk caused by the recently discovered M-dwarf companion. Thus, R~CrA may be a triple system hosting a disk.}
   {}

\begin{document} 
   \maketitle
%
%-------------------------------------------------------------------
\section{Introduction}
\label{sec:introduction}

%\subsection{Herbig Ae/Be stars}

Exoplanets form within protoplanetary disks around young stars (see e.g. \citealp{Chen2012, Marshall2014}).  Observations suggest that giant planets form more frequently around intermediate mass stars than around solar-mass stars (see e.g. \citealp{Johnson2010, Nielsen2019}). Herbig AeBe (HAeBe) stars \citep{Herbig1960, Hillenbrand1992} are still embedded in gas-dust envelopes and are frequently surrounded by circumstellar disks \citep{Perez1997}. They are young ($<$10 Myr), of intermediate mass (1.5-8$M_\odot$), with spectral type typically A and B (sometime F), and show strong Balmer emission lines in their spectra. They are characterized by a strong infrared excess \citep{The1994} due to a warm circumstellar disk, and represent the more massive counterparts of T~Tauri stars. They are experiencing accretion and their spectral line profiles are very complex.  Moreover, a quite consistent number of Herbig Ae/Be are  strongly variable stars \citep{vioque2018}, with typical periods from days to weeks \citep{Eiroa2002, Oudmaijer2001} and amplitudes of magnitudes in the optical.

%\subsection{The particular role of R~CrA}

The Coronae Australis (CrA) molecular cloud complex is one of the nearest region with ongoing star formation (distance $\sim$130~pc, \citealp{Neuhauser2008}). In its center there is the Coronet cluster characterized by a very high and variable extinction ($A_V$ up to 45 mag). R~CrA (HIP93449) is a HAeBe star and it was the first variable star identified in the CrA molecular cloud. Its spectral type is highly debated and estimates vary from F5 (e.g. \citealp{Hillenbrand1992}) and A5 (e.g. \citealp{Chen1997}), to B5III peculiar \citep{Gray2006})  and B8 (e.g. \citealp{Bibo1992}). %Its parallax from Gaia Data Release 2 (Gaia Collaboration 2018) is of 10.53$\pm$0.70 mas, corresponding to a distance of 94.2$\pm$6.7~pc, but likely this result is not correct because all other objects of the Coronet cluster are located at about 100~pc according to GAIA DR2.
R~CrA is much brighter in the near infrared (NIR) (J=6.94; H=4.95; K=3.46 mag, \citealp{Ducati2002, Cutri2003}) than in the visual (V=11.92 mag, \citealp{Koen2010}). Its brightness varies with time by up to 4 mags and its spectrum appears highly reddened. R~CrA was observed as part of the SHINE (SpHere INfrared survey for Exoplanets) survey \citep{Chauvin2017}, the main GTO program of SPHERE@VLT (Spectro-Polarimetric High-contrast Exoplanet REsearch, \citealp{Beuzit2019}). The images acquired with SPHERE revealed a quite complex system, which includes a previously unknown M-dwarf stellar companion (Mesa et al., 2019), together with a bright jet-like structure and the possible presence of a circumstellar disk outside the stellar companion. 
%These structures, together with an inner disk not imaged by the SPHERE observations, might be responsible of the highly variable brightness of the star.

\citet{Takami2003} proposed the presence of a stellar companion and of an outflow based on their spectro-astrometric observations at the Anglo-Australian Telescope. They estimated for the binary a separation of 8 au and a period of $\sim$24 years, values that are in disagreement with those obtained for the companion found in high contrast imaging. The presence of a companion was also proposed by \citet{Forbrich2006} in order to explain the X-ray spectrum of R~CrA as strong X-ray emission is not expected from HAeBe stars. 

R~CrA has been monitored by several photometric campaigns since about 120 years ago. \citet{Bellingham1980} detected long-period optical variation, while \citet{Herbst1999A} found a variability of \textbf{several tenths of mag} but no period shorter than 30 days on the Maidanak Observatory data. A period of 66~days was noticed on the AAVSO data by \citet{Percy2010}. 

The main goal of this paper is to investigate the origin of the variability of R~CrA by combining archive photometric and spectroscopic data. Since the fundamental physical quantities of the star derived in the literature are somewhat controversial, and such quantities are crucial for our study, we first derive them using a new methodology and the analysis of part of the spectral energy distribution (SED). The methodology is assessed using the photometric and spectroscopic data of a homogeneous sample of well studied HAeBe stars. In the following analysis we started analyzing the photometry of Herbig Ae/Be stars (Sect.~\ref{sec:phot}), setting the context. With the purpose of studying the R~CrA variability, we then made use of several archival data and Rapid Eye Mount (REM) telescope proprietary data. These are presented in Sect.~\ref{sec:data} and light curves are extracted in Sect.~\ref{sec:lightcurve}. A model that well matches observational data is described in Sect.~\ref{sec:model}. In Sect.~\ref{sec:spectrum}, we combine these photometric data with spectroscopic ones coming from directory discretionary time (DDT) observations obtained with SINFONI at ESO VLT telescope in 2018 \citep{Mesa2019}. Conclusions are drawn in Sect.~\ref{sec:conclusions}.
 
\section{SED and stellar parameters}
\label{sec:phot}
 
\begin{table*}[htb]
\centering
\caption{Basic data for Herbig Ae-Be stars in our sample; M$_{\rm V~W2}$ and  M$_{\rm V~abs}$ are the visual absolute magnitudes obtained from the W2 magnitude and from consideration of the interstellar absorption A$_V$}
%\tiny
%\small
\begin{tabular}{lccccccccccc}
\hline
Star            & $\pi$ & V  & J & W2 & $\log{T_{\rm M}}$& A$_V$ & M$_{\rm V~W2}$&  M$_{\rm V~abs}$ & $\log{L/L_\odot}$ &  Age & M \\
                & mas & mag & mag & mag & & mag & mag & mag & & Myr & M$_\odot$ \\
\hline
\hline
\multicolumn{12}{c}{Objects with spectra dominated by emission lines} \\
\hline
R~CrA          &6.54&11.51&~6.935&1.377&3.98&5.47&-0.30&~0.14&2.12&1.5& 3.5 \\
V380~Ori       &2.86&10.90&~8.107&3.411&3.97&2.93&~0.03&~0.02&2.02&2.4& 3.1 \\
MWC~1080       &0.73&11.85&~7.460&1.666&4.48&5.52&-4.78&-3.12&4.70&1.0&18.7 \\
HD~250550      &1.40&~9.59&~8.475&4.633&4.02&0.49&-0.21&-0.63&2.27&1.4& 3.7 \\
HD~259431      &1.38&~8.72&~7.454&3.169&4.15&1.06&-1.72&-2.06&3.12&1.0& 6.5 \\
LkHalpha~348   &0.75&14.30&~9.962&2.880&3.80&6.71&-3.54&-0.88&2.84&1.0& 5.4 \\
LkHalpha~198   &1.67&13.79&~9.990&4.123&4.02&5.03&-0.51&~0.21&2.16&1.7& 3.5 \\
PV~Cep         &2.91&17.46&12.453&3.315&4.00&9.34&-0.27&     &2.17&1.5&3.6 \\
R~Mon          &1.25&11.85&~9.686&1.337&4.08&5.84&-3.95&-0.11&3.03&1.0& 6.2 \\
BD+61~154      &1.76&10.60&~8.137&3.579&4.08&2.48&-0.84&-0.94&2.58&1.1& 4.2 \\
BD+40~4124     &1.09&10.62&~7.904&2.879&4.27&3.17&-2.60&-2.26&3.63&1.0& 9.2 \\
LkHalpha~134   &1.18&11.42&~9.457&6.569&4.03&0.39&~1.37&-0.38&1.92&3.9& 3.0 \\
LkHalpha~233   &1.14&13.52&11.290&4.926&3.93&4.00&-0.50&~0.29&2.01&1.8& 3.2 \\
\hline
\multicolumn{12}{c}{Intermediate Objects} \\
\hline
AB~Aur         &6.14&~7.05&~5.936&2.142&3.97&0.44&~0.52&~0.09&1.90&3.1& 3.0\\
LkHalpha~234   &0.80&12.73&~9.528&3.546&4.11&4.56&-2.66&-1.63&3.14&1.0& 6.6\\
HD~150193      &4.62&~8.79&~6.947&3.244&3.95&1.06&~0.98&~0.24&1.75&3.4& 2.9\\
HD~163296      &9.85&~6.85&~6.195&2.466&3.97&0.00&~1.89&~1.55&1.33&9.2& 2.0\\
HD~200775      &2.77&~7.36&~6.611&2.220&4.27&0.66&-1.14&-1.28&3.14&1.0& 6.6\\
BD+46~3471     &1.29&10.15&~8.540&4.532&3.98&1.13&-0.51&-0.91&2.32&1.1& 4.0\\
Z~CMa          &4.30&~8.80&~6.543&1.717&4.48&2.54&-0.76&~0.52&3.16&2.0& 2.0\\
\hline
\multicolumn{12}{c}{Objects with spectra dominated by absorption lines} \\
\hline
BS5999/V856~Sco&6.14&~7.05&~5.907&1.895&3.89&0.68&~0.26&~0.31&1.83&2.3& 3.0\\
KK~Oph         &4.22&10.99&~9.072&3.179&3.88&3.24&~0.63&~2.40&1.33&6.2& 1.5\\
HD~97048       &5.41&~9.00&~7.267&4.300&4.02&0.24&~2.41&~0.83&1.45&10.3& 2.3\\
HK~Ori         &2.17&11.10&~9.408&4.850&3.93&1.73&~0.92&~1.18&1.55&4.2& 2.1\\
LkHalpha~215   &1.40&10.70&~8.598&4.840&4.15&1.36&-0.04&-0.87&2.56&1.6& 4.4\\
XY~Per         &2.14&~9.80&~7.654&3.786&3.91&1.51&-0.17&-0.71&2.13&1.2& 3.8\\
VV~Ser         &2.38&11.80&~8.673&3.914&4.14&3.31&~0.12&-0.05&2.34&1.9& 3.6\\
LkHalpha~208   &1.49&11.30&10.254&6.613&3.89&0.23&~1.92&~1.62&1.23&7.2& 1.6\\
\hline
\end{tabular}
\normalsize
\label{tab:similar}
\end{table*}

\begin{figure}[ht]
    \centering
    \includegraphics[width=\columnwidth]{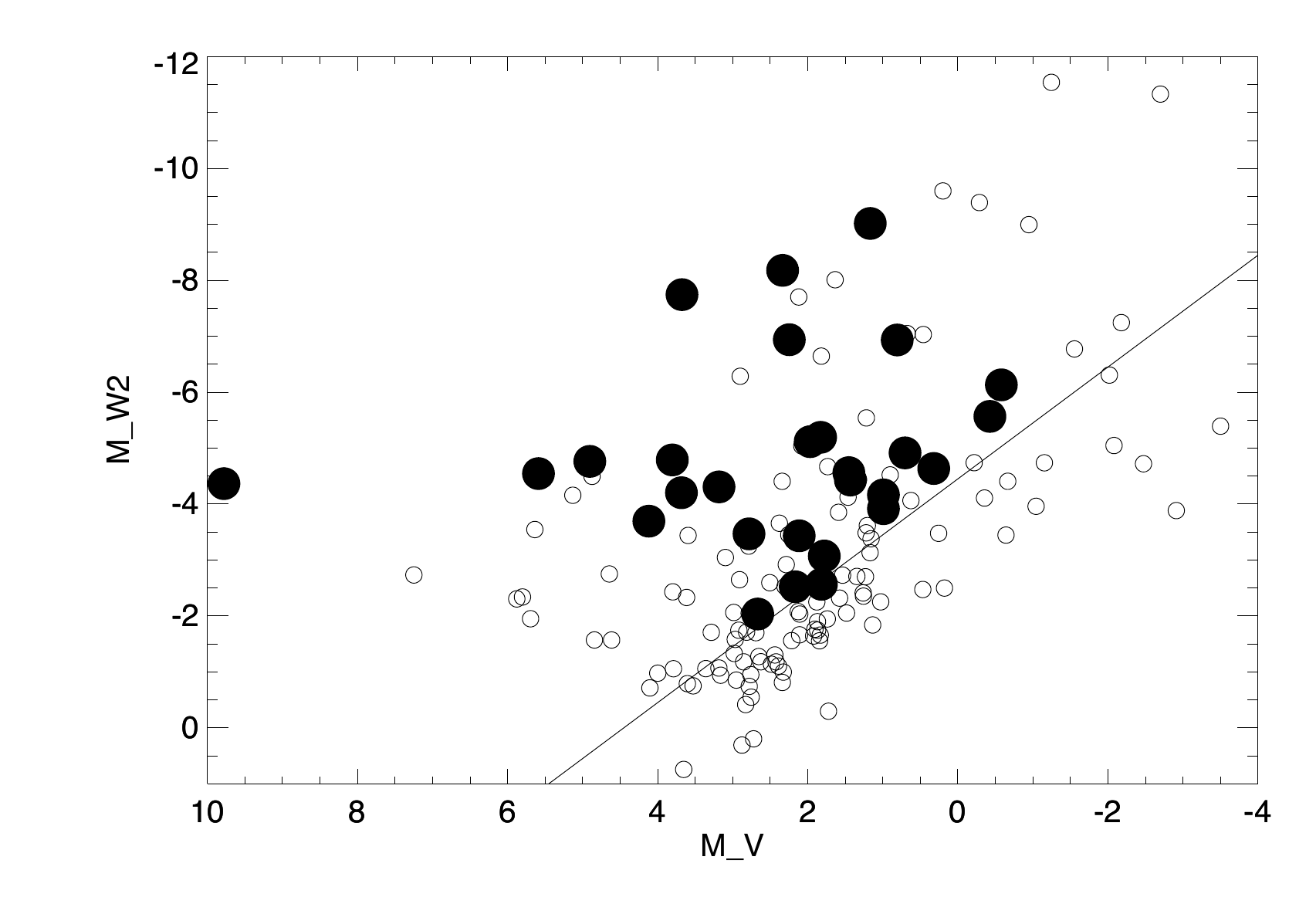}
    \caption{Comparison between the absolute magnitude in the V band and in the W2 band for samples of Herbig Ae-Be stars from \citet{Hamann1992} (filled circles) and \citet{vioque2018} (open circles); neither of them are corrected for absorption. The solid line is the relation $M_{\rm V}=M_{\rm W2}+4.45$ that we assume to be representative of unreddened objects}
    \label{fig:mv_w2}
\end{figure}

\begin{figure}[htb]
    \centering
   \includegraphics[width=\columnwidth]{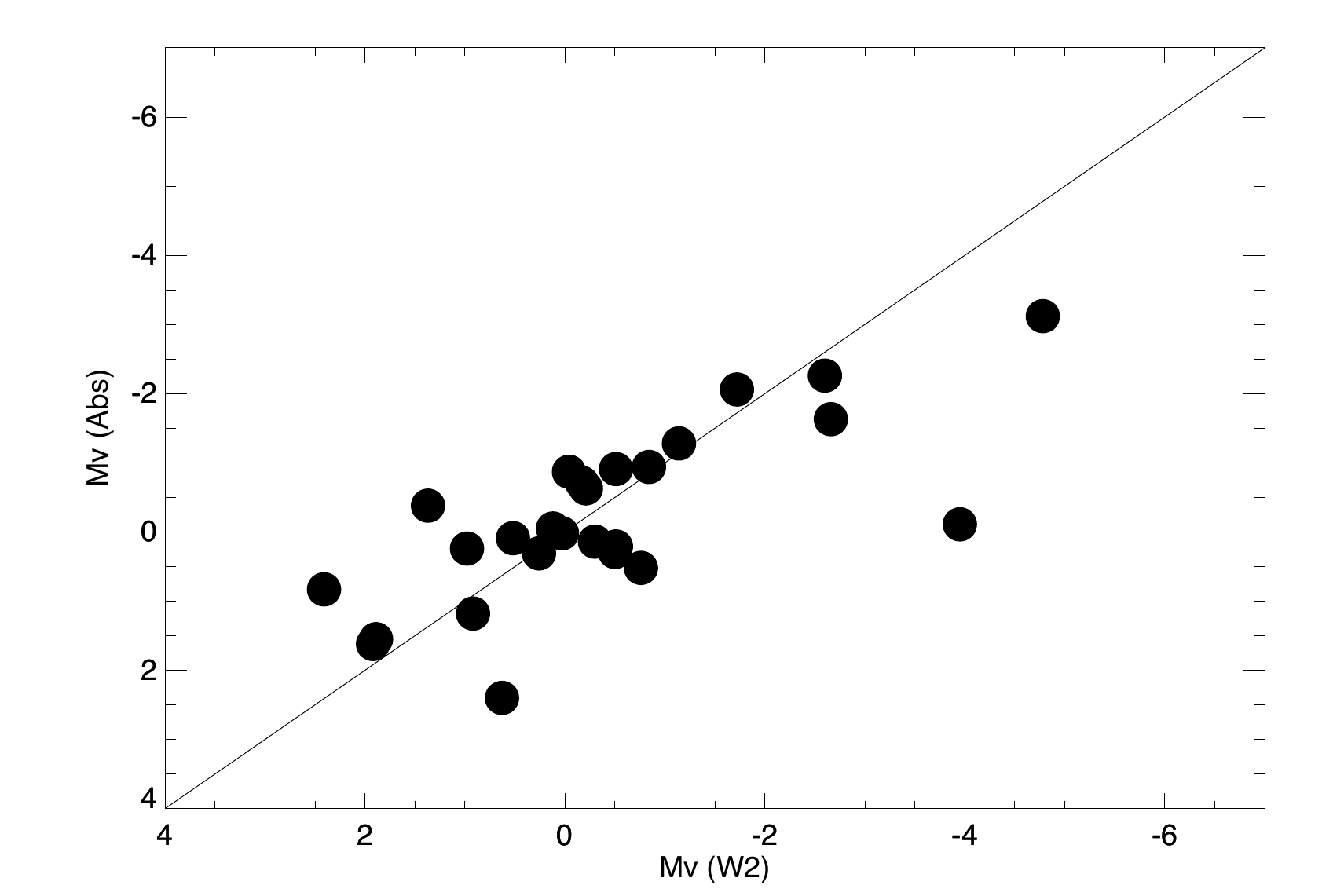}
    \caption{Comparison between absolute magnitude $M_{\rm  V}$ obtained from W2 and from correction for absorption for a sample of Herbig Ae-Be stars. The solid line represents equality between the two determinations}
    \label{fig:mv}
\end{figure}

The SED of HAeBe stars, in particular that of \rcra, is very complex because of the contribution of different components to the total emission, and their relative weight at different wavelengths. Even limiting us to the UV-NIR portion of the spectrum, we may outline four main components \citep{Merin2004, Meijer2008}:
\begin{itemize}
    \item the photospheric emission, possibly reddened by the absorption of circumstellar and interstellar dust; this should dominate at optical wavelengths
    \item a hot spot on the star surface corresponding to the shock in the accretion region; this should dominate in the ultraviolet
    \item the warmer part of the disk, dominating in the near IR (above 2~$\mu$m), with possibly some contribution by the cool part of the disk in the thermal infrared. Note that dust absorption is not large in this spectral region
    \item possibly, some light reflected by the circumstellar matter; this is essentially a scaled photospheric spectrum.
\end{itemize}
Above 2~$\mu$m, we expect that the emission from the warmest part of the disk is dominant. In the case of R~CrA this is shown not only by the shape of the SED, but also by the interferometric observation obtained with VLTI/AMBER \citep{Kraus2009}. This resolved the source into two Gaussians, with FWHM of about 25~mas (4 au) and 5.8~mas (0.9 au), with the second one providing about 2/3 of the total flux in the K-band. The prevalence of the disk emission in the NIR has been later confirmed by \citet{Lazareff2017} using PIONEER observations. In general, if the disk extends inwards up to the sublimation limit of the dust (see also \citealp{Lazareff2017}), it reaches temperature in the range 1500-1800~K, whatever the emission of the central source is. This is not possible if for instance a companion at a distances of the order of an au breaks the disk at short separation. \citet{Lazareff2017} estimated a dust sublimation temperature of 1650~K, based on the analysis of the SED of several HAeBe stars. They also argue that this is consistent with an inner rim temperature of 1800~K. However, they also noted that the disk should have a range of temperatures. In this work, we adopt a critical temperature of 1500~K; indeed we will consider longer wavelengths than considered in their study, with a stronger contribution by regions at lower temperatures. In agreement with \citet{Lazareff2017}, this implies a close similarity between the SED of the inner part of the disk among different objects. Namely, the emission of this region of the disk should be roughly proportional to the luminosity of the star - provided the disk flaring is not too much different from star-to-star. In turn, the luminosity should be proportional to the fourth power of the temperature, as long as the stellar radius does not change too much among HAeBe stars. As we will see, even if these assumptions appear very rough, they allow constraining the temperature and other characteristics of the stars.

We may then assume that the absolute luminosity in the WISE W2 band (central wavelength 4.6~$\mu m$, pass-band width of 1.1~$\mu m$) is a good proxy for the total luminosity for most of the HAeBe stars. This to be true, we should assume that the thickness of the disk, and then the fraction of stellar light intercepted by the dust and re-processed into thermal radiation, is the same for all the stars. For those cases where absorption is negligible (not a correct assumption in general), the V-W2 color yields directly an estimate of the correction to the W2 absolute magnitude needed to obtain the absolute magnitude of the star. In this way, one can obtain the luminosity of the star from the V magnitude, and the stellar mass by comparison with pre-main sequence isochrones, given a suitable temperature. If the temperature is also known, for instance from the spectral type, then age and stellar mass can be immediately derived.

The most critical step in this process is the derivation of the correction to the W2 absolute magnitude needed to obtain the absolute magnitude of the star. This is not an easy task for individual stars, especially in the presence of absorption, as in the case of \rcra, where we use the typical nir-IR SED for an HAeBe star. We first selected a sample of such stars and applied this analysis to all of them. We then looked for those with the lowest absorption i.e. those star with the lowest value of the V-W2 colour and whose disk is likely to be seen face on - to extract the proper correction to the W2 magnitude. In this work, we exclude those cases that do not have a warm disk at all.

We used two compilations of HAeBe stars: the extensive one by \citet{vioque2018}, that however does not include R~CrA, and a more limited sample by \citet{Hamann1992}, that includes R~CrA (see Table~\ref{tab:similar}). The parallax for the stars are from GAIA DR2 \citep{Gaia2018}, except for R~CrA itself, for which we adopt the parallax estimated in \citet{Mesa2019}\footnote{As noticed by \citet{Mesa2019}, the GAIA DR2 parallax for R~CrA is far from that measured for all other members of the Coronet cluster, for which consistent values were obtained.}. The V magnitude is from SIMBAD catalogue, the J magnitude from 2MASS \citep{Skrutskie2006}, and the W2 one from the WISE catalogue \citep{Cutri2013}. With this approach we find that $M_{\rm V}=M_{\rm W2}+4.45$\ for those stars at the lower envelope of the distribution (solid line in Fig.~\ref{fig:mv_w2}).  Once the zero-points of V and W2 photometry are considered, we end up with the result that the warm disk re-converts in thermal emission a fraction equal to 0.17 of the total energy emitted by the star.  Assuming that the disk has an albedo equal to zero, this corresponds to an optically thick inner disk, up to an effective angle of $\pm 9.5$\textdegree from the disk equator (this angle corresponds to a sky area equal to 17\% of total). This value roughly agrees with the value typically expected for inner disks around HAeBe stars from interferometric measures \citep{Lazareff2017}. It is worth noting that in the case of accreting T~Tauri stars, the typical disk-to-stellar luminosity ratio ($L_{\rm disk}/ L_{\rm star}$), drawn from their SED by combining optical, 2MASS and Spitzer data, is $\sim$20\%--30\% (see histograms in Fig.~14 of \citealp{Merin2008} and Fig.~12 of \citealp{Alcala2008}). Using data provided in those works we indeed found a strong linear correlation between $L_{\rm disk}$ and $L_{\rm star}$, and confirm that $L_{\rm disk}/L_{\rm star} \approx 0.25$. This correlation may be attributed partly to reprocessing of the stellar light by the dusty disk, but also to the correlation between stellar luminosity, and hence stellar mass, with the accretion luminosity (see \citealp{Alcala2017}). A similar correlation for HAeBe stars is thus expected. \citet{Kenyon1987} defined the regime of accreting disks as those with $L_{\rm disk}/L_{\rm star}>$0.1, while for passive reprocessing disks $0.02 < L_{\rm disk}/L_{\rm star} < 0.08$. According to this classification, \rcra\ (that has $L_{\rm disk}/L_{\rm star}=0.17$) has an accreting disk.
 
Once the correction needed to convert the W2 magnitude into the V one is known, we considered the stars in the sample by \citet{Hamann1992}. We divided these stars into three groups, according to their spectra: (1) the spectrum is dominated by emission lines; (2) intermediate cases; (3) the spectrum is dominated by absorption lines. We excluded the two stars (HD~52721 and HD~53367) that do not have a near infrared excess detectable in the WISE photometry. The values of absolute $V$-band magnitude  M$_{\rm V~W2}$ estimated with this procedure are listed in Table~\ref{tab:similar}.

We may assess our methodology by comparing these estimates of the stellar luminosity M$_{\rm V~W2}$ with those obtained by a different approach where we use the apparent V magnitudes and correct them for an estimate of the extinction (hereafter M$_{\rm V~abs}$). This may be derived from a comparison of the observed colours with those expected for pre-main sequence stars with the same effective temperature. To this purpose, we should only use the portion of the SED dominated by the stellar photosphere: in practice, we selected the V-J colour because the accretion contribution is large at wavelengths shorter than V, and we wish to have a wide enough spectral range for this purpose. The contribution of the disk in the J-band is not at all negligible\footnote{In principle, the V-R or V-I colour may better represent the emission from the photosphere \citep{Bessell1988}, avoiding contamination by the disk emission in the IR. However, we considered here the radiation re-processed by the disk in the J-band.} that should be considered in the analysis. In our schematic model - where the temperature of the inner disk is fixed at 1500~K and the fraction of the stellar luminosity reprocessed through the disk is constant - the correction for the disk contribution only depends on the temperature of the star, and is well represented by the following equation in the temperature range $6000< T_{\rm eff}< 30000$~K:
\begin{equation}
dJ=-6.36\times 10^{-10}~T_{\rm eff}^2-7.50\times 10^{-5}~T_{\rm eff}+0.34 
\end{equation}
where $dJ$\ is indeed the contribution of the disk in the J-band. This is derived by fitting data obtained by combining a hot black body (of variable temperature) describing the photospheric spectrum with a cooler one (with temperature of 1500~K) that re-process 17\% of the radiation. We considered here the stellar temperatures determined by \cite{Manoj2005b}, and compared the observed colours with the relation by \citet{Pecaut2013} for pre-main sequence stars. We then converted the $E(V-J)$\ reddening values into $A_{\rm V}$ values using the reddening law by \citet{Cardelli1989} with $R_{\rm V}=4.7$. As discussed by \citet{Lazareff2017}, this value of $R_{\rm V}$\ is in the high range of those found in the literature and should be more appropriate for highly extincted HAeBe stars \citep{Hernandez2004}, while lower values are found for HAeBe stars with low extinction \citep{Montesinos2009, Fairlamb2015, Blondel2006}\footnote{Conversely, the value of $R_{\rm V}$\ might be determined by comparing the observed colours of the stars with those of suitable templates. As an exercise, we considered the multi-band photometry of R~CrA by \citet{Koen2010}, limiting to the BVRI$_C$\ data because photometry in the U-band may be affected by accretion. We then assumed that the colours of the star are the same as those of pre-main sequence stars with similar $T_{\rm eff}$ \citep{Pecaut2013}. With these assumptions, we find that best agreement between observed and template colours is obtained for $R_{\rm V}=4.7$, confirming the value adopted in our analysis. This large value of $R_{\rm V}$ suggests that absorption towards R~CrA is due to grains with typical size larger than generally considered for the interstellar medium.}. We give emphasis in this paper to the first class of objects, that includes R~CrA. These are also those objects where the precise value of $R_{\rm V}$\ is more important.

Figure~\ref{fig:mv} compares the absolute luminosity obtained using these two different approaches for the stars listed in Table~\ref{tab:similar}. The agreement is good; the Pearson linear correlation coefficients are $r=0.92$, 0.98 and 0.63 for $A_{\rm V}$, age, and mass, respectively, that all are highly significant for this sample of 20 stars. In particular, the values we obtain for R~CrA are $M_{\rm V~W2}=-0.30$ using the W2 approach, and $M_{\rm V~abs}=0.14$\ using the absorption one. Averaging the two results and taking into account the bolometric correction using the table by \citet{Pecaut2013}, we obtain an absolute luminosity of $\log{L/L_\odot}=2.12$.

The most obvious outlier in this figure is R~Mon, for which we derive a much larger luminosity from W2 photometry than from the extinction correction method. In addition, also KK~Oph appears discrepant. Since all these objects are variable and we are using photometry taken at different epochs, some scatter in this comparison is expected. For instance, the V magnitude used for KK~Oph (V=10.99) is much brighter than other values listed in the literature (V=12.36: \citealp{vioque2018}; V=11.45: \citealp{Herbst1999A}). Adopting the Vioque et al. value, the absolute visual magnitude would be $M_{\rm V}=0.74$, in better agreement with the value obtained from the W2 photometry.

\begin{figure}[htb]
\centering
\includegraphics[width=\columnwidth]{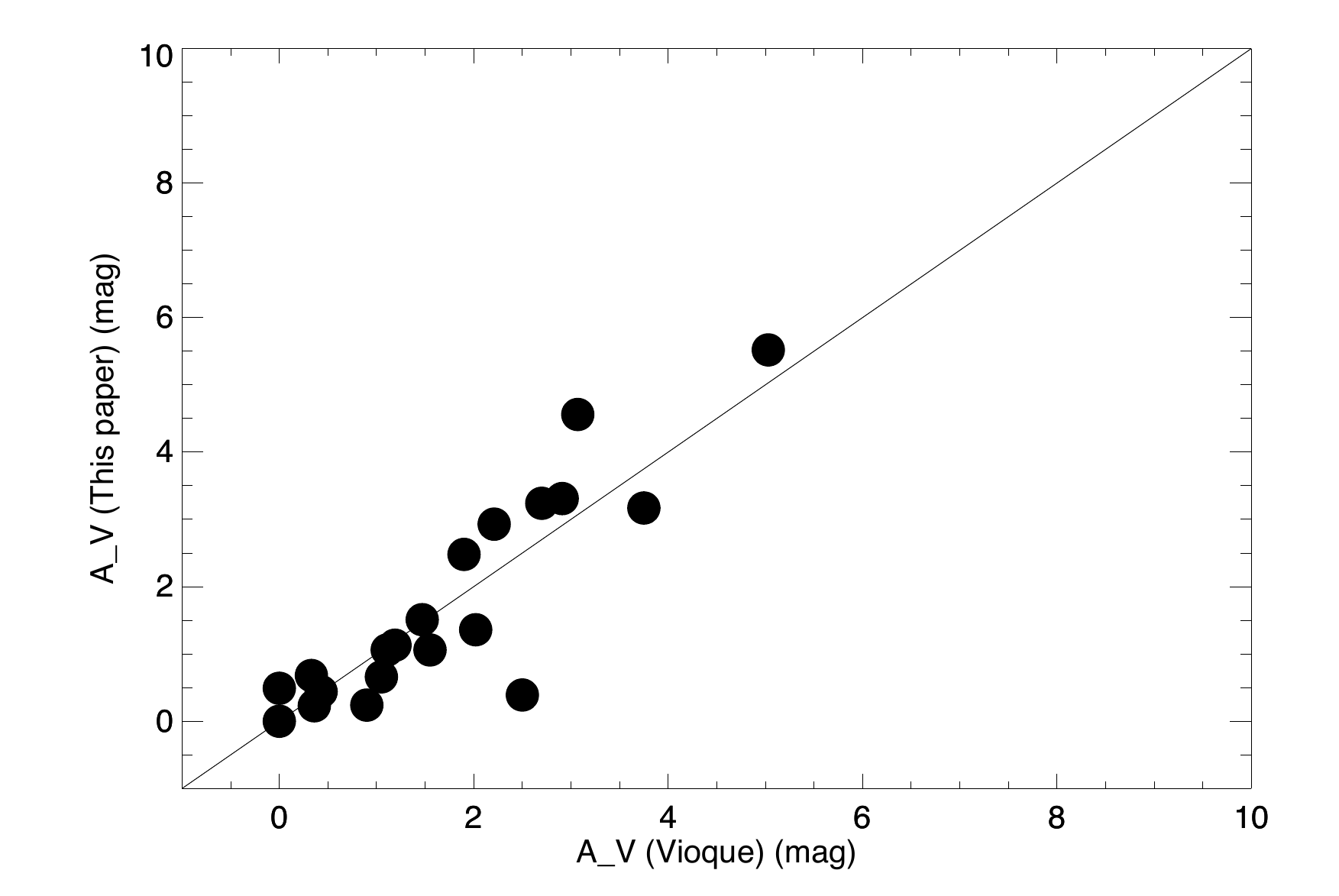}
\includegraphics[width=\columnwidth]{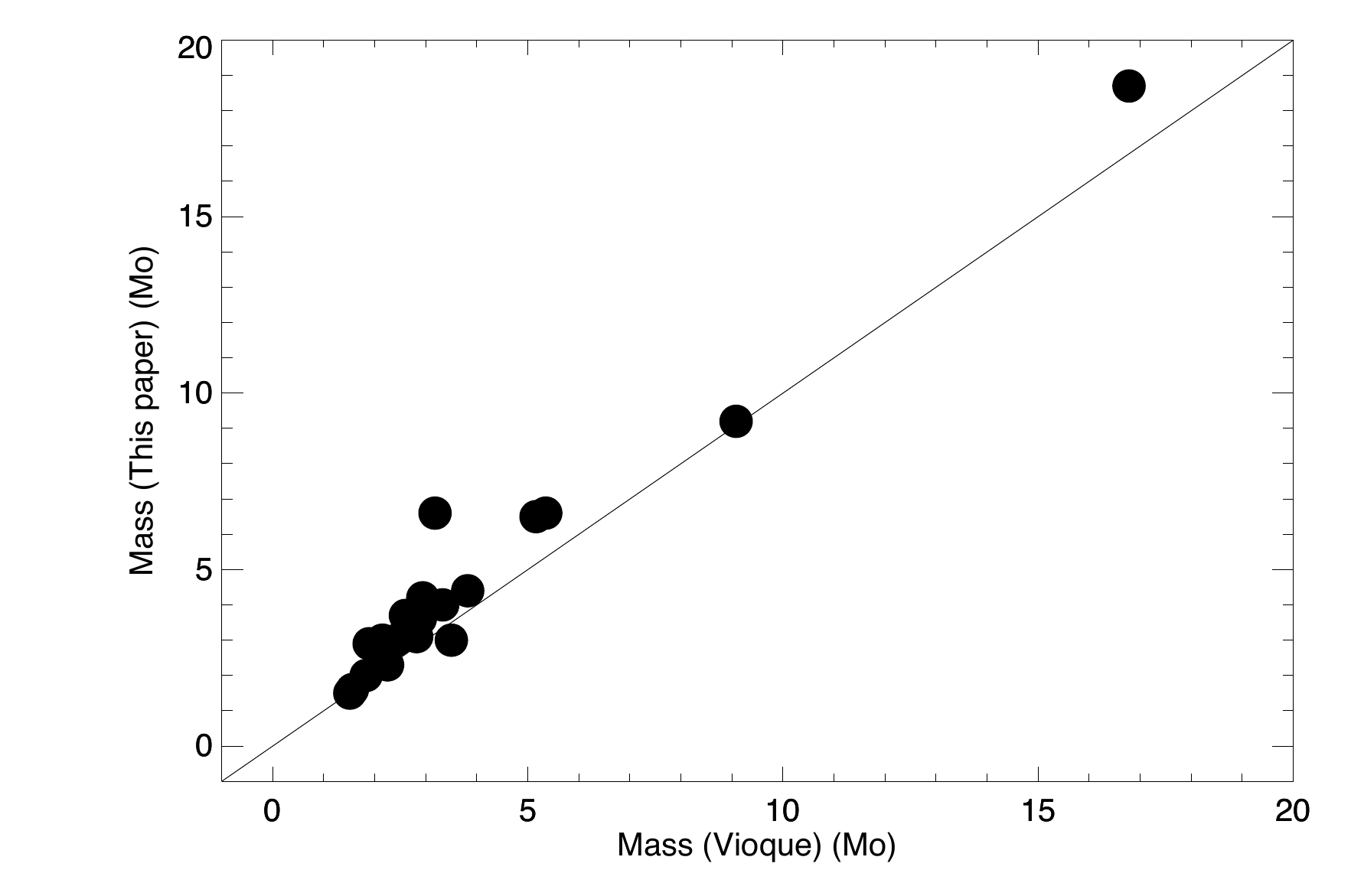}
\includegraphics[width=\columnwidth]{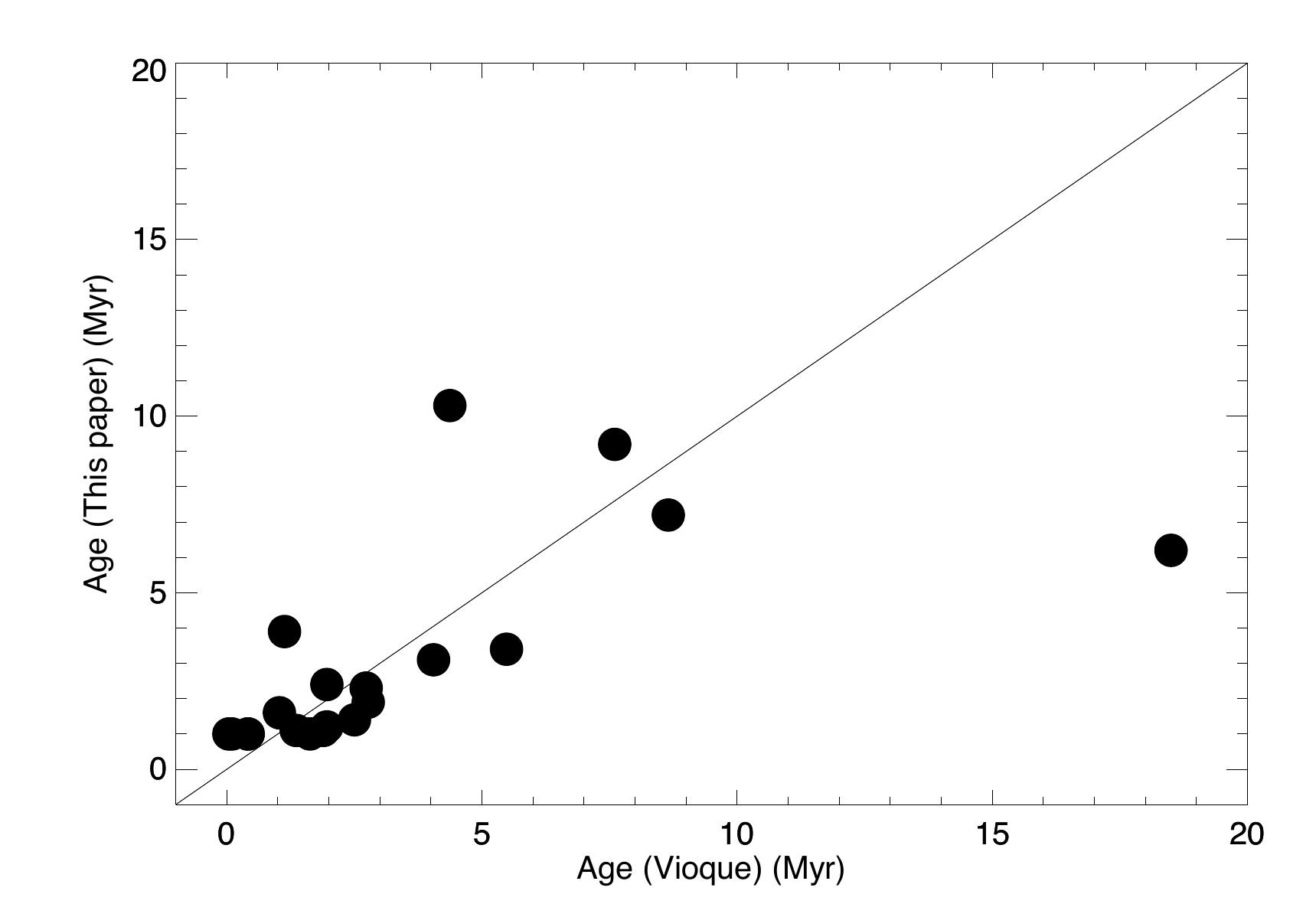}
\caption{Comparison between the values derived in this paper and those from \citet{vioque2018}.  Upper left panel: absorption in the V band A$_V$; upper right panel: mass; lower panel: age. The solid line represents identity.}
\label{fig:vioque}
\end{figure}

We may compare this luminosity with that predicted by the pre-main sequence models by Pisa group \citep{Tognelli2011}; entering the temperature (from \citealp{Manoj2005b}) we may then extract an age and a mass for the stars. These values are listed in the last two columns of Table~\ref{tab:similar}. In order to further assess the reliability of these determinations, in Fig.~\ref{fig:vioque} we compared our values of absorption in the V-band, mass and age with those listed by \citet{vioque2018}. The agreement is generally good (save for the ages of KK~Oph and HD~97048) and supports the current estimates. 

Error bars estimation is not straightforward since they are dominated by uncertainties in the methods rather than in the observable. We derive a reasonable guess by comparing the results obtained using different approaches. On average, differences are of about 0.5 mag in $A_V$, 15\% in the masses, and of 1.5~Myr in the ages. We will adopt these values as error bars, being actually a prudent approach. The values we obtain for R~CrA are then $1.5\pm 1.5$~Myr and M=$3.5\pm 0.5$~M$_\odot$ for the age and mass, respectively. We notice that this approach assumes that R~CrA is a single star; we will see later that it is most likely a binary with two quite similar components and will revisit these values accordingly. The absorption in the visual band we obtain ($A_V=5.47\pm 0.5$~mag) is slightly larger than the value obtained by \citet{Bibo1992} ($A_V=4.65$~mag).

\rcra\ belongs to the group of HAeBe stars with emission line dominated spectra. To this group belong stars that are, on average, younger (the median value is 1.5 Myr, with a root mean square scatter - rms - of individual values of 0.8 Myr, to be compared with 2.0 Myr - rms of 2.9 Myr - for the second group and 3.2 Myr - rms of 3.3 Myr - for the third one), more massive (median value of 3.7~M$_\odot$ - rms of 4.2~M$_\odot$, with the second group at 3.0~M$_\odot$ - rms of 2.0~M$_\odot$ -, and the third one at 2.7~M$_\odot$ - rms of 1.1~M$_\odot$), and more absorbed (median value of 4.0~mag - rms of 2.1~mag -, with the second group at 1.4~mag - rms of 1.3 mag -, and the third one at 1.8~mag - rms of 1.0  mag). The age and the mass of these stars are not surprising: the strength of the emission lines is in fact likely related to accretion that is expected to decline with age and mass. Noteworthy, the higher extinction might be attributed, in part, to a larger average distance. However, the correlation between distance and absorption shows a considerable scatter. 

Among the stars with spectra dominated by absorption lines, R~CrA is by far the one nearest to the Sun. AB~Aur and BS5999/V856~Sco are at a similar distance, but their spectrum is less extreme than that of R~CrA, and in fact these stars are older and less massive. R~CrA is also very strongly absorbed, though there are more extreme cases on this respect in this sample. A summary of the properties of R~CrA is given in Table~\ref{tab:my_label}.

%--------------------------------------------------------------------

\begin{table}[htb]
\caption{Stellar parameters}
    \centering
    \begin{tabular}{lcc}
    \hline
    \hline
    Parameter & Value & Ref.  \\
 \hline
Distance & 150 pc & 1\\
Luminosity & $\log{L/L_\odot}=2.46$ & 2 \\
Temperature &  $\log{T_{\rm eff}}=3.98$~K&   2\\
Radius (if a single star): & 6.2 $R_\odot$& 2\\
Mass (1 Myr isochrone): & 4.13 $M_\odot$& 2\\
\hline
    \end{tabular}
    \tablebib{(1)~\citet{Mesa2019}; (2)~\citet{Manoj2005b}
}
\label{tab:my_label}
\end{table}

\section{Photometric data}
\label{sec:data}

\begin{table}[htb]
    \centering
     \caption{Summary of archival data used in this work}
    \begin{tabular}{lcccc}
    \hline\hline
    Archive & Start & End & n$_{\rm obs}$ & Band \\
    \hline
AAVSO      & 1896 & 2017 & 9927 & Vis \\
MAO        & 1983 & 1992 &  240 & UBVRI \\
ASAS       & 2001 & 2009 & 1437 & Vis \\
SuperWASP  & 2006 & 2008 & 9139 & Vis \\
ASAS-SN    & 2014 & 2018 &  206 & Vis \\
REM        & 2018 & 2018 &   23 & g'r'i'z'JHK \\
\hline
\end{tabular}
\label{tab:catalogs}
\end{table}

R~CrA is a variable star and the correct interpretation of the light curves at different wavelengths is crucial to understand its nature. Table~\ref{tab:catalogs} lists a number of photometric data sets for R~CrA. In the following, we give some details about them.

\subsection{AAVSO}

The American Association of Variable Star Observer (AAVSO) is a community of amateur and professional astronomers that manages an archive of more than 34 million variable star observations. \citet{Percy2010} studied the photometric stability of R~CrA exploiting the 100-year long sequence provided by AAVSO in the visible band including about 10,000 measures and more than 1,000 upper limits. They found a period of 66 days, stable in time but with slightly variable peak-to-peak photometric amplitude and that the observed V magnitudes vary from 10.5 to 14.5. 
 
\subsection{MAO}

A homogeneous photometric survey of R~CrA was carried out at visible wavelengths at the Maidanak Observatory in Uzbekistan starting from 1983, and was published in 1999 by Herbst \& Shevchenko. %\cite{Herbst1999A}.

\subsection{ASAS}

The All Sky Automated Survey (ASAS; \citealp{Pojmanski1997}) is a project started in 1997 to monitor the luminosity of all stars brighter than 14 magnitudes exploiting two observing stations one in Las Campanas Observatory, Chile (operating since 1997) and the other on Haleakala, Maui (operating since 2006). Both are equipped with two wide-field 200/2.8 instruments, observing simultaneously in V and I band. However, only V data are available in the on-line catalog. We found 1437 aperture photometry points for R~CrA between 2001 and 2009. 

\subsection{ASAS-SN}

The All-Sky Automated Survey for Supernovae (ASAS-SN; \citealp{Shappee2014, Kochanek2017}) currently consists of 24 small-diameter telescopes, distributed around the globe that allow covering the whole visible sky every night. This survey provides 206 photometric point of R~CrA in the visible spanning from 2014 to 2018.

\subsection{SuperWASP}

The Wide Angle Search for Planets (SuperWASP: \citealp{Butters2010}) uses two telescopes at Roque de los Muchachos (La Palma) and the South African Astronomical Observatory. Each telescope uses an array of eight 200 mm f1.8 Canon lenses feeding a 2k$\times $2k CCD. Data of R~CrA cover the epochs from 2006 to 2008. These observations were acquired with a broad-band filter covering a pass-band from 400 to 700 nm.

\subsection{REM}

We observed \rcra with the Rapid Eye Mount (REM: \citealt{Chincarini2003}) telescope located in La Silla. REM provides simultaneous photometry in the optical and NIR using the ROSS2 and REMIR instruments, observing in the g', r', i', z', and JHK band filters, respectively. The observations started in the night of August 7, 2018 and lasted till September 30, 2018 for a total of 23 points for each observing band. 
% The bright star HD~176386A ($\alpha:$19 01 39 $\delta:$-36 53 27, V=7.326 mag, H=46.809 mag, separation of  $\sim$310 arcsec from R~CrA, that is within the ROSS and REMIR field of view) was used as photometric calibrator, since its luminosity is stable with respect to the variations we are interested in. 
Observations allowed us to follow the photometric variation at seven different bands. The extension to the near infrared is of particular relevance to study the variability of R~CrA.

\section{Light curve discussion}
\label{sec:lightcurve}

\subsection{Secular evolution}

\begin{figure}[htb]
    \centering
    \includegraphics[width=\columnwidth]{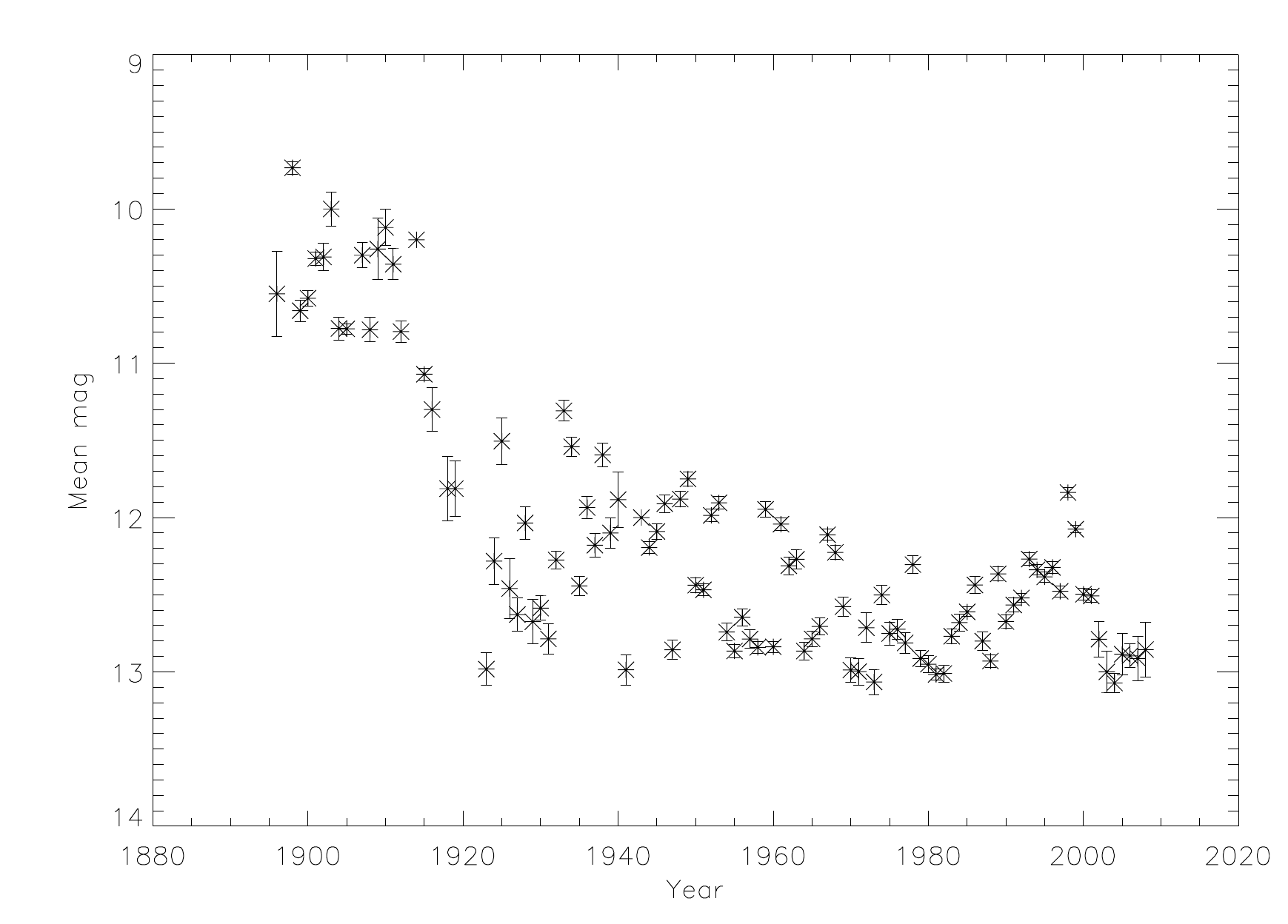}
    \caption{Secular variation of AVVSO photometric measurements. Median values for each year are shown.}
    \label{fig:AAVSO_sec}
\end{figure}

The apparent magnitude of R~CrA evolved significantly in the last century. Figure \ref{fig:AAVSO_sec} shows the secular evolution of the magnitude of R~CrA in the visual band since 1890, obtained using the AAVSO photometry. We plotted the median for each year. R~CrA had a magnitude of $\sim 10$ before 1915; it then declined by about two magnitudes between 1915 and 1920, and then oscillated between magnitude 12 and 13. This secular variation might either be an indication of a decrease in the accretion rate or of a variation on the dust absorption. We will come back on this point in Sect.~\ref{sec:model}. On top of this secular evolution, R~CrA shows a short term variability. In the following part of this section, the annual median magnitude was subtracted from each photometric point in order to discuss the variability on shorter timescales.

\subsection{Periods/phased curves}

\begin{figure}[htb]
    \centering
\includegraphics[width=\columnwidth]{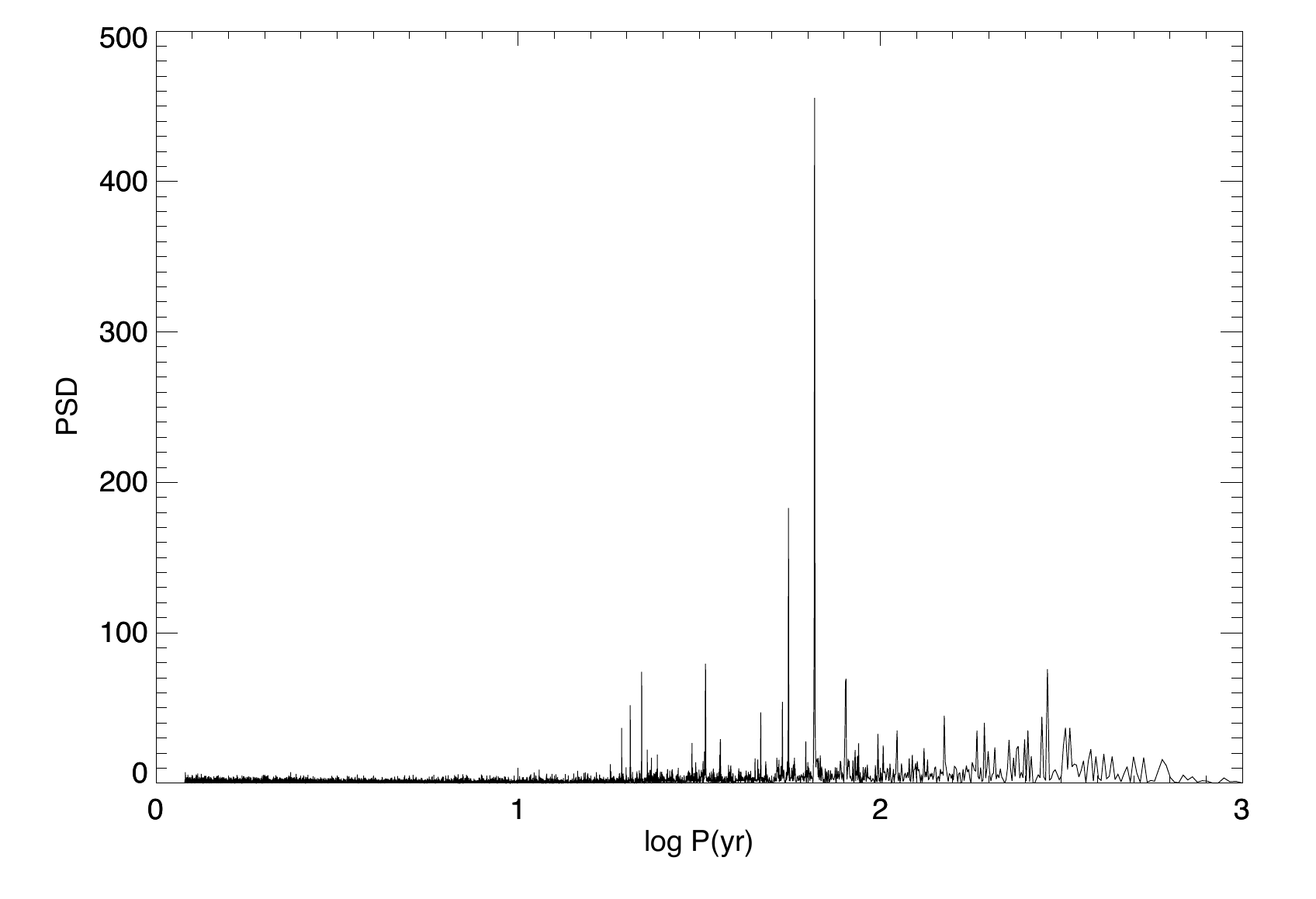} 
\includegraphics[width=\columnwidth]{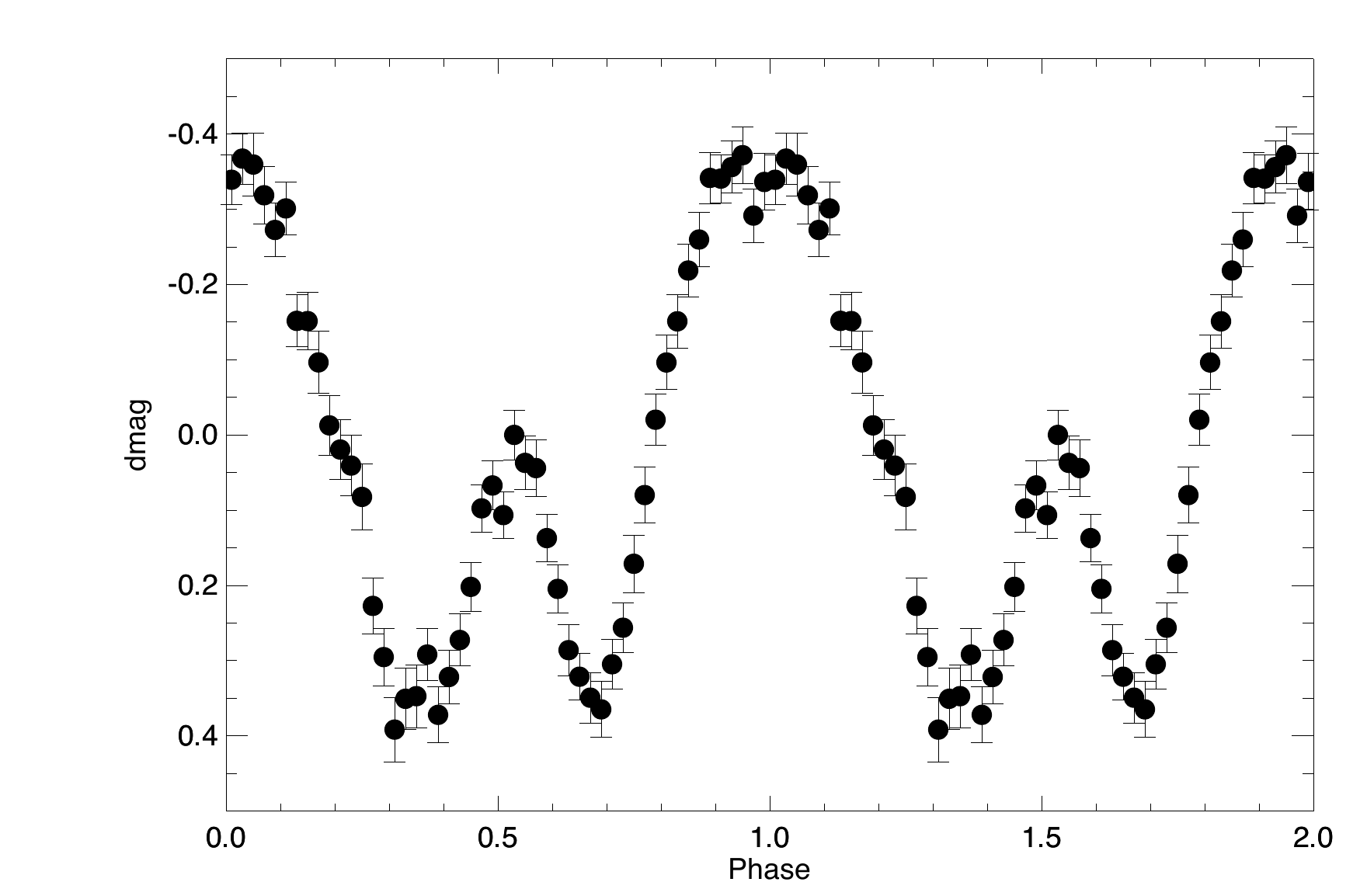}
    \caption{Top panel: Generalized Lomb Scargle periodogram for the whole series of AAVSO photometric data for R~CrA, once the median for every year has been subtracted; Bottom panel: light curve phased at the peak of the periodogram (65.767~days). Points are the average for each phase bin and error bar are the standard deviation of the mean.}
    \label{fig:aavso}
\end{figure}

%\begin{figure*}
%    \centering
%\includegraphics[width=\columnwidth]{magvvstime.eps} 
%\includegraphics[width=\columnwidth]{RCrA_RV_phased.eps}\\
%\includegraphics[width=\columnwidth]{periodogramma_frequenza_zoom.eps} 
%\includegraphics[width=\columnwidth]{window_function_zoom.eps}\\
%    \caption{Upper left: photometric data in V band obtained by \citet{Herbst1999A}; upper right: The data rephased at 66 days; lower left: Generalized Lomb Scargle periodogram for R~CrA photometric data; lower right: The related Window Function.}
%    \label{fig:herbst}
%\end{figure*}

 As shown by several authors \citep[see e.g.][]{Bellingham1980, Percy2010}, most of the short term variability of R~CrA can be described as a periodic behaviour. This is obvious from Fig.~\ref{fig:aavso} that shows the Lomb-Scargle periodogram (GLSP, \citealp{Zechmeister2009}) of the AAVSO data (from which we removed the secular trend as discussed above). The peak of this diagram is for a period of $65.767\pm 0.007$~days, in agreement with previous estimate by \citet{Percy2010}. The other relevant peaks correspond to the first and second harmonics of this period and aliases at 1 year of the fundamental mode together with the two first harmonics. The relative power of the harmonics shows that the shape of the light curve is not reproduced by a single sine wave. 
%A similar result is obtained using e.g. the \citet{Herbst1999A} data (see Fig. \ref{fig:herbst}). In this case the GLSP shows several significant peaks, with separations in frequency corresponding to one year (0.00276~day$^{-1}$). The two highest power frequency are those of 55.7 days and 65.9 days, the second one corresponding to the period found in the AAVSO data and the first being its alias to one year.

Once phased at the best period derived from the GLSP, the shape of the light curve of R~CrA becomes obvious (see lower panel of Fig.~\ref{fig:aavso}). To reduce the impact of other source of variations - e.g. accretion - and exploiting the richness of the photometric data series, in this figure (and in the similar ones shown in the Appendix) we plotted the mean values at each 0.02 phase bin, with the error bar given by the standard deviation of these mean values. The light curve is almost symmetric, with two maximums of unequal brightness. As shown in the Appendix, a similar shape with only minor variations is obtained also considering various subsets of the whole AAVSO data series (e.g. considering ranges 10 years long) as well as all other photometric data series obtained at visible wavelengths. This light curve is then consistently found over about 120 years of observations. A similar regularity reminds its origin to some basic property of R~CrA. Since the period appears to be long for stellar rotation, the most likely explanation is a binary orbital period (see Sect.~\ref{sec:model}).
 
\subsection{Variation with colour}

\begin{figure}[htb]
    \centering
   \includegraphics[width=\columnwidth]{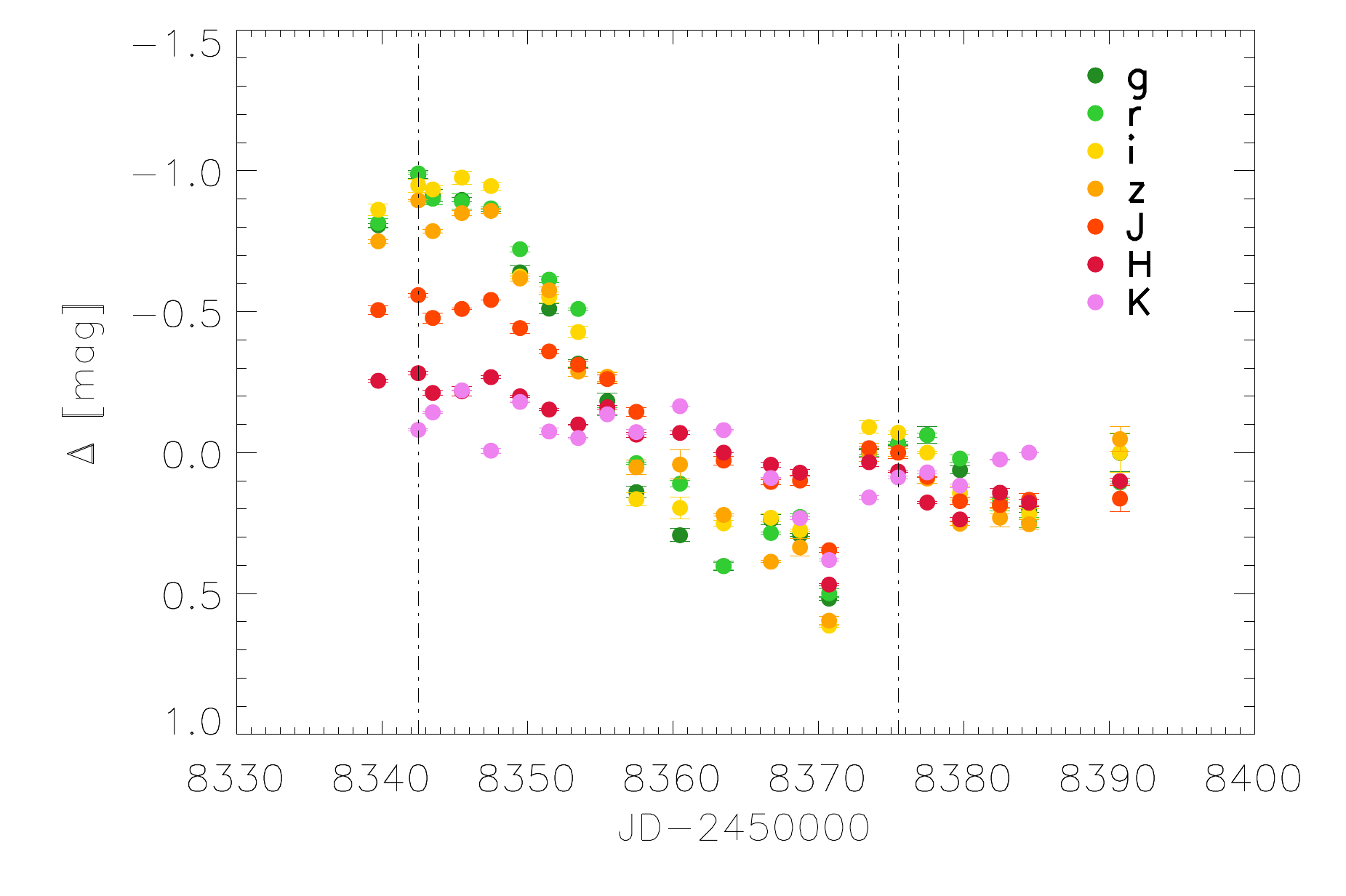}
\includegraphics[width=\columnwidth]{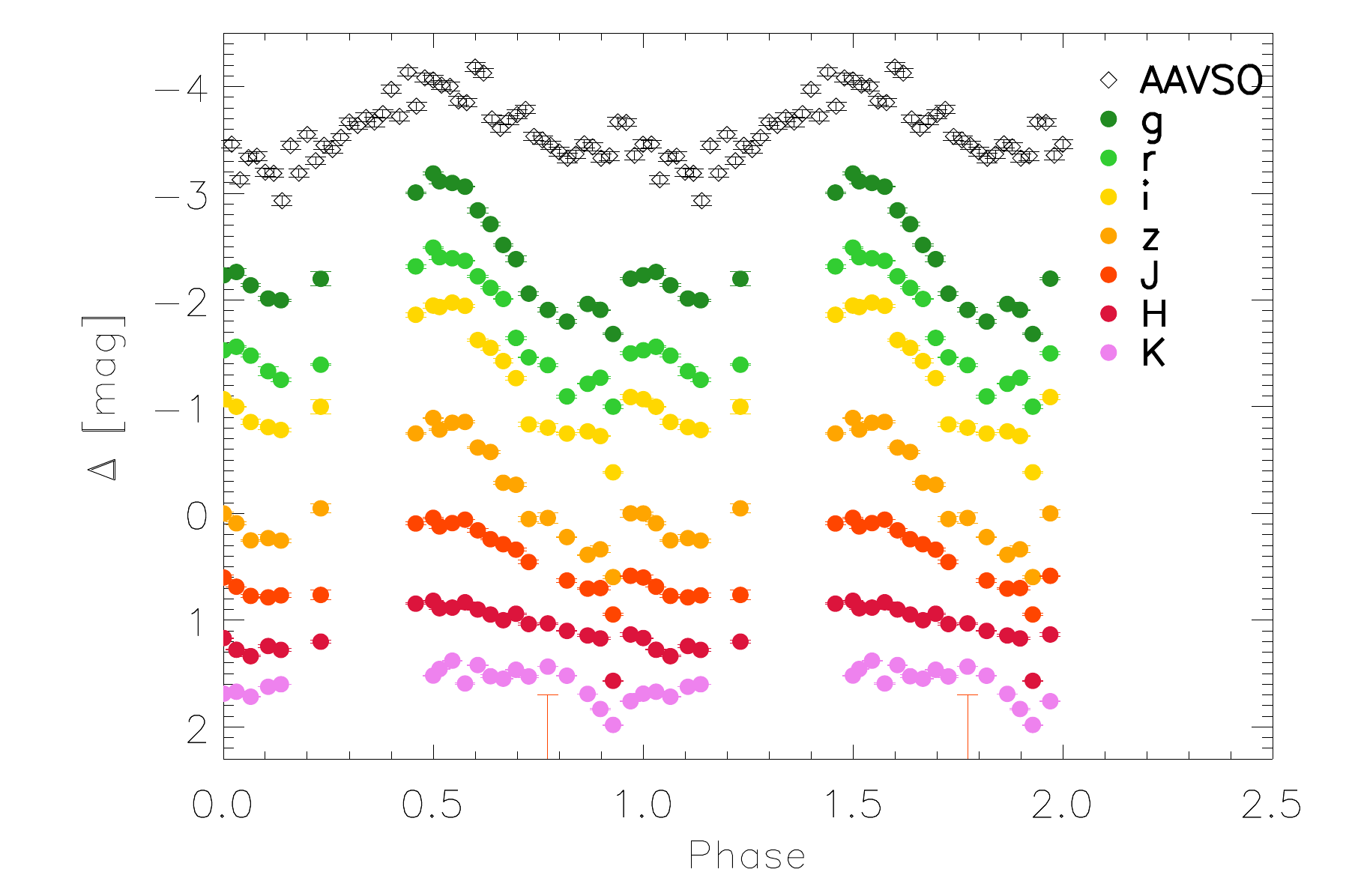}\\
    \caption{Top: photometric variation of R~CrA from REM images. Bottom: same as top but phased to the 65.767 day period according to the AAVSO sequence (black). For better visualization, we applied an offset to the zero point of each sequence.}
    \label{fig:rem}
\end{figure}

The large wavelength coverage provided by the REM data allows a comparison of the light curves obtained in the visible and near infrared (NIR). As for other Herbig AeBe stars, we expect that at short wavelengths (for instance, $g'$) luminosity is mainly due to accretion, while in the NIR (for instance, K band) is mainly due to the hottest part of the circumstellar disk. At intermediate wavelengths (e.g. $i'$), we expect the contribution of the photosphere to be dominant. $r'$ may be influenced by emission in $H\alpha$. In addition, the wide spectral coverage of REM may help to model the (possibly variable) dust absorption by the disk itself, that is seen at an inclination of $>70$ \textdegree (see \citealt{Mesa2019}).

With the aperture photometry method we were able to trace the variability at all wavelengths for 80\% of the phase, as shown in Fig. \ref{fig:rem}. In the bottom panel of Fig. \ref{fig:rem}, we compare the phased AAVSO sequence with the REM one, but zero points of the magnitude were varied to allow a better visualization of the results. As expected, the light curve at optical wavelengths is indeed very similar to that found in AAVSO observations; the amplitude reduces progressively with wavelength in the NIR, with a very small variation (if any) in the K-band.

\section{Modelling the light curve}
\label{sec:model}

\begin{figure}[htb]
\centering
\includegraphics[width=\columnwidth]{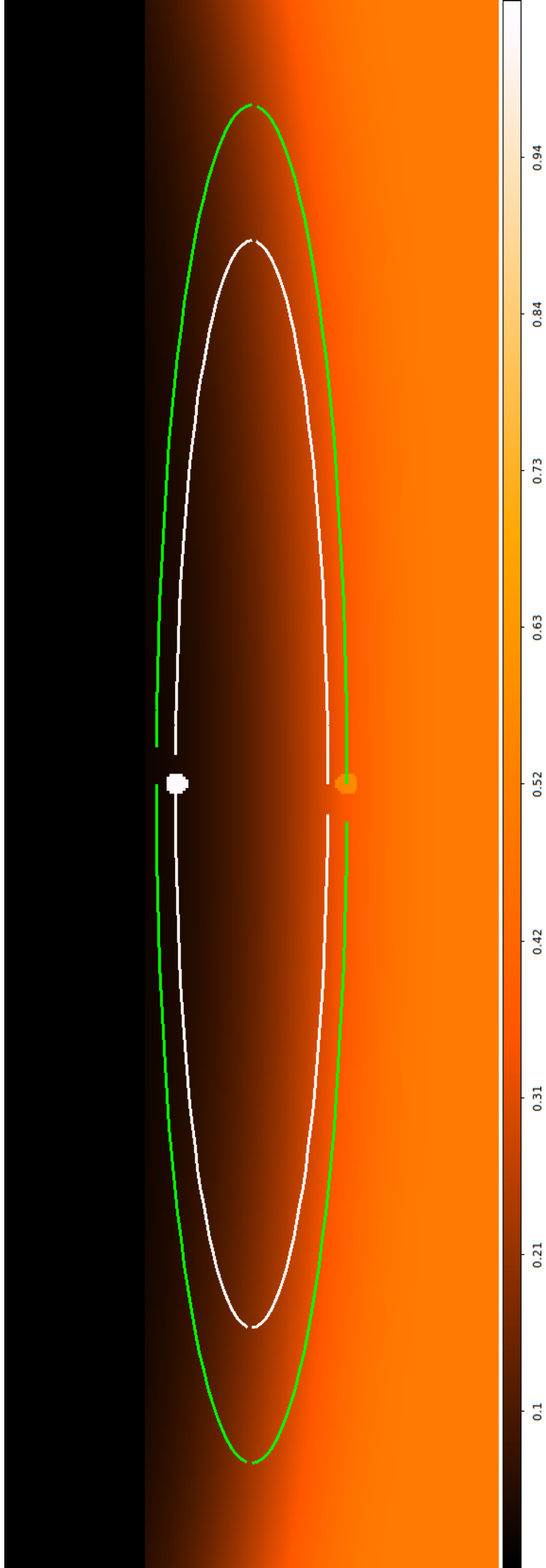}
\caption{Schematic view of the extinction model for R~CrA. The white and green ellipses represent the apparent orbits of the two components in the plane of the sky. The reddish region represents the area covered by the near side of the disk. When one of the two components is in this region, its light is attenuated by the dust in the disk}
\label{fig:model}
\end{figure}

\begin{figure}[htb]
\centering
\includegraphics[width=\columnwidth]{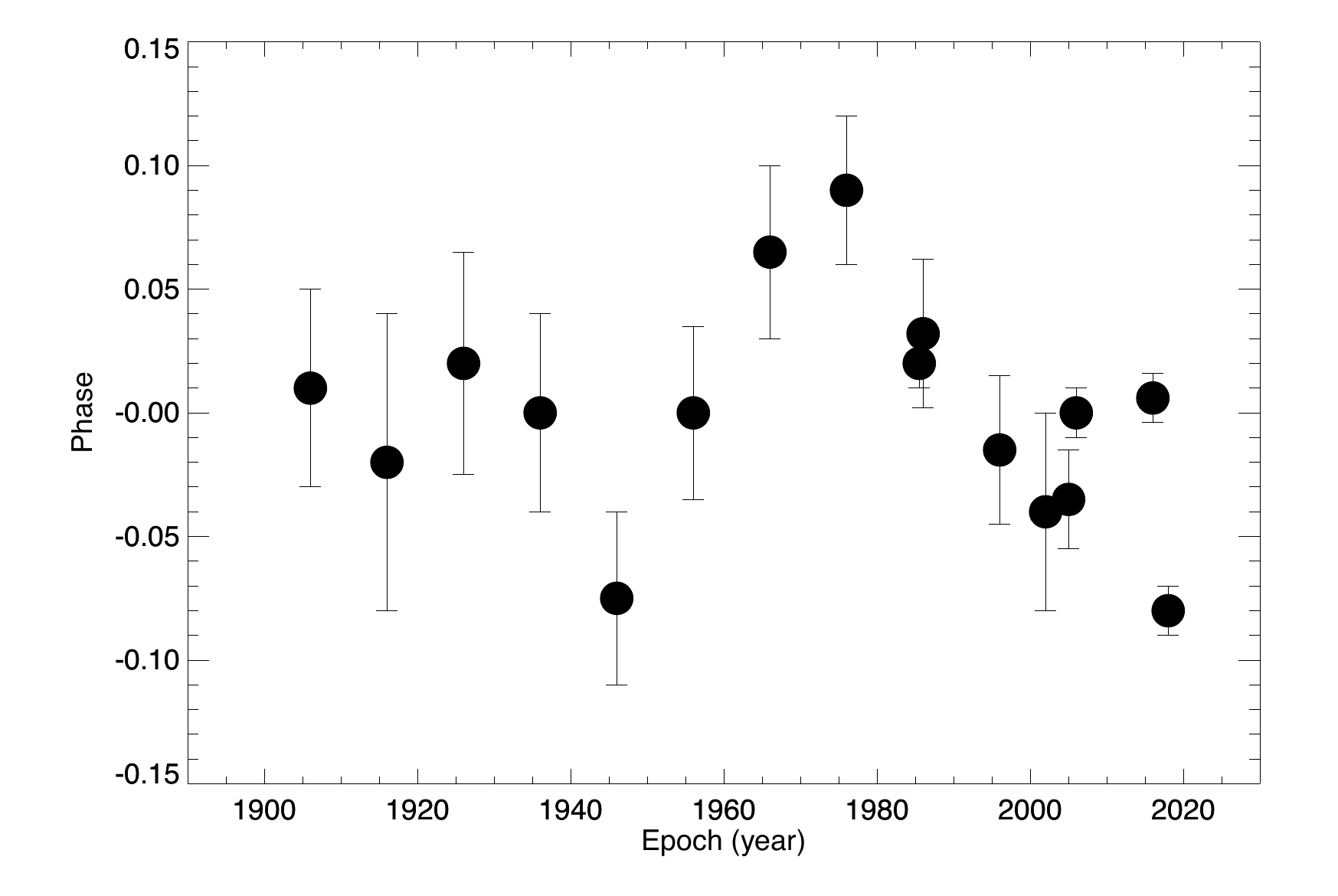}
\includegraphics[width=\columnwidth]{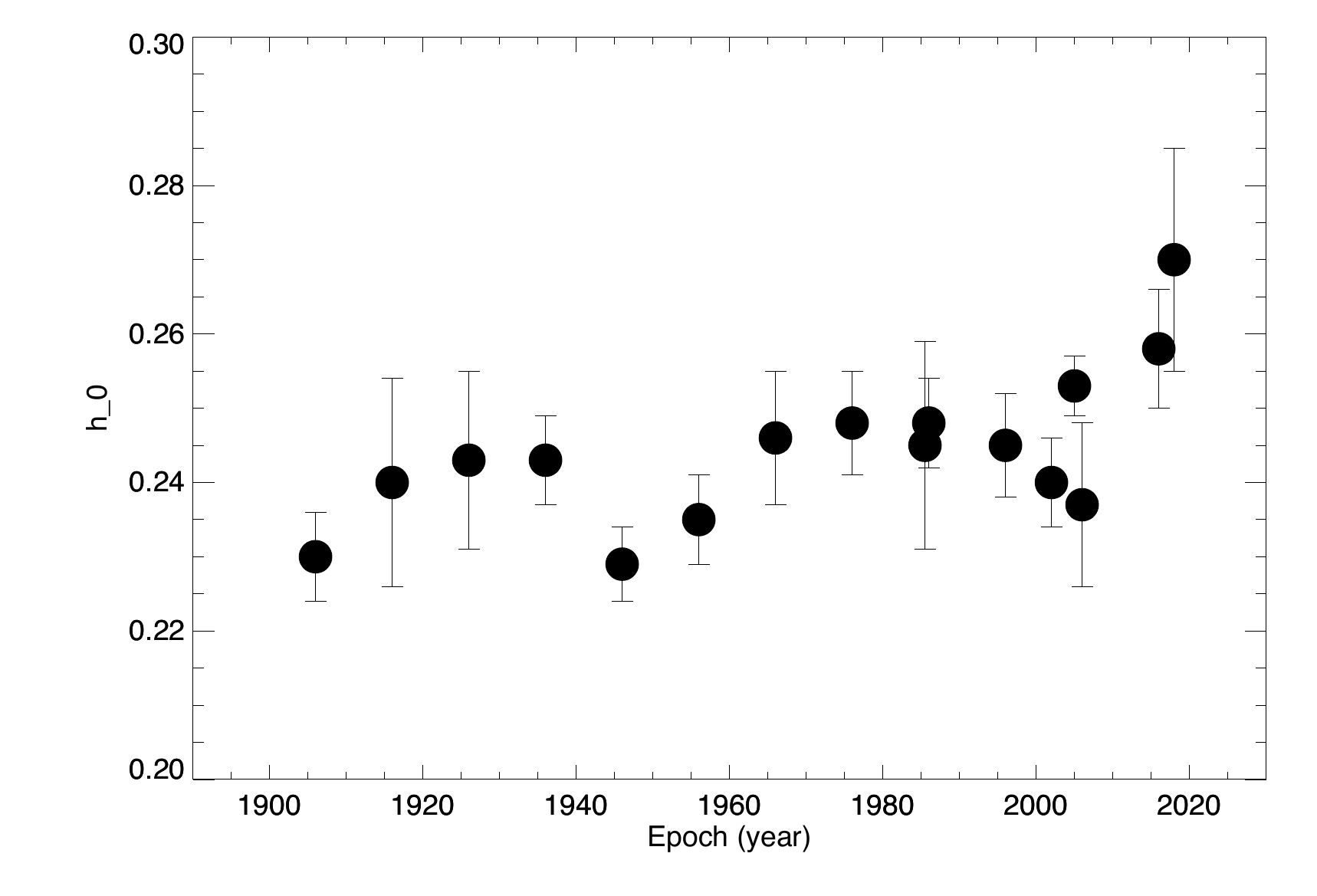}
\includegraphics[width=\columnwidth]{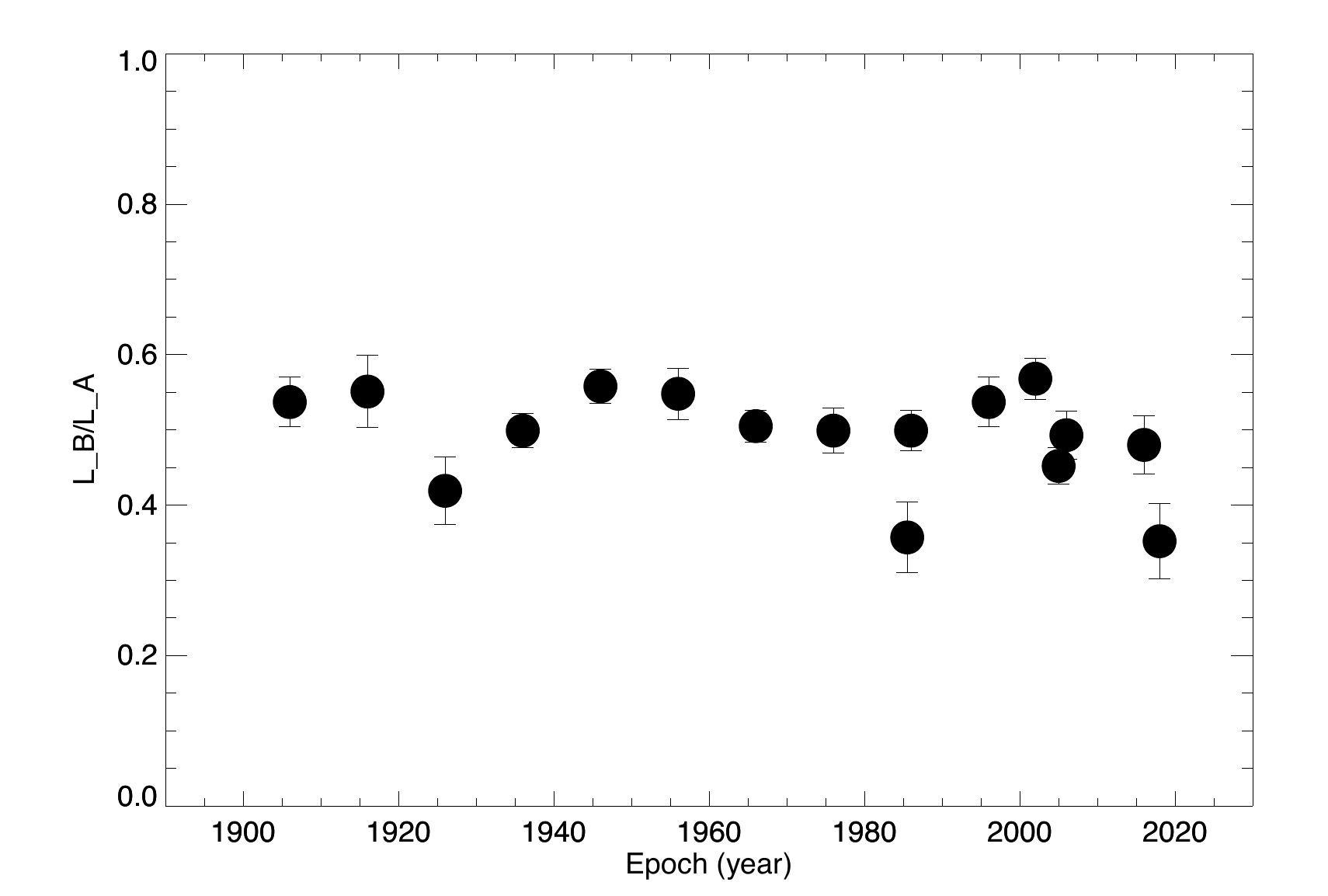}
\caption{Secular changes of the light curve parameters. Top panel: phase; middle panel: disk extinction height scale $h_0$; lower panel: luminosity ratio between the two components.}
\label{fig:model_par}
\end{figure}

\begin{table*}[htb]
\caption{Light curve parameters derived at different epochs}
\centering
\begin{tabular}{lcccc}
\hline
Source & Mean Epoch & $L_B/L_A$ & $h_0$ & Delay \\
\hline
AAVSO \#1 & 1906 & 0.537$\pm$0.033 & 0.230$\pm$0.006 &~0.010$\pm$0.04 \\
AAVSO \#2 & 1916 & 0.551$\pm$0.048 & 0.240$\pm$0.014 &-0.020$\pm$0.06 \\
AAVSO \#3 & 1926 & 0.419$\pm$0.045 & 0.243$\pm$0.012 &~0.020$\pm$0.045\\
AAVSO \#4 & 1936 & 0.499$\pm$0.023 & 0.243$\pm$0.006 &~0.000$\pm$0.04 \\
AAVSO \#5 & 1946 & 0.558$\pm$0.023 & 0.229$\pm$0.005 &-0.075$\pm$0.035\\
AAVSO \#6 & 1956 & 0.548$\pm$0.034 & 0.235$\pm$0.006 &~0.000$\pm$0.035\\
AAVSO \#7 & 1966 & 0.505$\pm$0.021 & 0.246$\pm$0.009 &~0.065$\pm$0.035\\
AAVSO \#8 & 1976 & 0.499$\pm$0.030 & 0.248$\pm$0.007 &~0.090$\pm$0.03 \\
AAVSO \#9 & 1986 & 0.499$\pm$0.027 & 0.248$\pm$0.006 &~0.032$\pm$0.03 \\
AAVSO \#10& 1996 & 0.537$\pm$0.033 & 0.245$\pm$0.007 &-0.015$\pm$0.03 \\
AAVSO \#11& 2002 & 0.568$\pm$0.027 & 0.240$\pm$0.006 &-0.040$\pm$0.04 \\
Herbst et al&1985.5&0.357$\pm$0.047& 0.245$\pm$0.014 &~0.020$\pm$0.01 \\
ASAS      & 2005 & 0.452$\pm$0.024 & 0.253$\pm$0.004 &-0.035$\pm$0.02 \\
SuperWasp & 2006 & 0.493$\pm$0.032 & 0.237$\pm$0.011 &~0.000$\pm$0.01 \\
ASAS-SN   & 2016 & 0.480$\pm$0.039 & 0.258$\pm$0.008 &~0.006$\pm$0.01 \\
REM       & 2018 & 0.352$\pm$0.05  & 0.270$\pm$0.015 &-0.080$\pm$0.01 \\
\hline
\end{tabular}
\label{tab:light_curve_param}
\end{table*}

Given the quite long period, it appears difficult to explain the light curve of R~CrA as due to either rotation or pulsations in single star. The shape of the light curve is also different from that expected due to mutual eclipses in a close binary system. We think the light curve of R~CrA can be interpreted using a model where the star is actually a rather close binary - with period equal to the fundamental period that can be extracted from the periodogram - surrounded by a circumbinary dusty disk. A similar scenario has been proposed to explain the variability of a class of extreme asymptotic giant branch (AGB) stars, the so-called RVb stars. Another case of partial attenuation by a circumbinary disk is that of the T Tauri star KH~15D (V582~Mon), see \citet{Aronow2018} and references therein\footnote{Also in that case, there is a very stable period of 48.37 days - that is thought to be the binary period - on top of a secular brightness evolution of a few mag. However, once phased to the binary period, the light curve of KH~15D is quite distinct from that of R~CrA, with single broad maximums and minimums, rather than unequal peaks as in R~CrA. This calls for a different geometry of the system.}. This scenario itself divides into two main alternatives, where the origin of the variation is related to a variable extinction (see e.g. the case of HR~4049 considered by \citealp{Waelkens1991}) or a variation of the light scattered by the disk (as proposed by \citealp{Waelkens1996} for the case of the Red Rectangle). Crucial to separate between these two alternatives is the wavelength dependence of the variability: in fact, if the amplitude of the light variation is fairly independent of wavelength, the contribution by the disk must be considered. Our REM data set indicates that the variability of R~CrA is indeed very different in the optical and NIR: this supports the extinction scenario, that we will adopt hereinafter to explain the light curve at optical wavelengths. On the other hand, the contribution by the disk is dominating in the NIR.

The RVb star model consists of two stars with large difference in the luminosity so that only the brightest component is observed. On the other hand, in the case of \rcra\ the light curve has two maximums, each one attributed to a different component; these have quite similar mass, with a luminosity ratio equal to the ratio between the luminosity of the primary and secondary peaks of the light curve.

In the RVb-stars scenario, the system is seen at an angle that is quite similar to the flaring angle of the disk, so that the star is strongly extincted by the disk when is close to conjunction, and much less when it is close to opposition. Within \rcra\ scenario, the two components are then partially hidden by the disk during their motion around the common center of gravity, the extinction depending on their exact location along the orbit (see sketch in Fig.~\ref{fig:model}). We may then construct a model that describes the light-curve considering that the stellar flux is due to two point-sources in Keplerian motion around a common center of gravity seen from a suitable angle, and that light from the two sources is behind an absorbing slab that has a variable extinction as a function of the height over the orbital mid-plane. Binary orbit and disk are here assumed to be co-planar. As mentioned above, the mass ratio between the two components can be obtained from the luminosity ratio between the two maximums, using a reasonable mass-luminosity ratio: we found that this implies a mass ratio close to 0.77. We further assume that the disk extinction may be represented by a Gaussian functional form such as $\exp{-(z/h_0)^2}$, where $h_0$\ represents half the disk thickness assuming that the absorbing slab is located at a distance from the common center of gravity equal to the binary semi-major axis. Within these assumptions, there is a degeneracy between the inclination of the disk and $h_0$. To remove this degeneracy, we assumed that the system is seen at an inclination of 80\textdegree; while this is an arbitrary assumption, it allows to have a disk flaring angle as defined by $h_0$ that is $\sim 15$\textdegree, a value that is quite reasonable for a circumstellar disk around a Herbig Ae-Be star. A lower inclination would imply a larger flaring angle. We notice that a too large inclination angle ($i>86$\textdegree) can be excluded by the absence of eclipses. 

Within this model, the amplitude (and shape) of the light curve is very sensitive to the exact value we assume for $h_0$, that is then accurately determined from observations. In fact, if $h_0$\ is very small, the light curve is characterized by a nearly constant value with only two - unequal - minimums: this is because the system is seen at a moderate inclination, so that the disk hides each of the two stars only when they are close to conjunction. When the disk hides the stars even at quadrature, the light curve assumes the typical shape with two unequal maximums that we observe in the case of R~CrA; however, the depth of the minimums in the light curve that occurs when both stars are close to quadrature increases rapidly with increasing $h_0$. Note that this curve is symmetric if we assume that the orbit is circular and that the disk extinction is independent of the azimuth angle.

We used our model to fit separately the light curves obtained at different epochs and from different sources (see Appendix). In the case of the AAVSO data, that extends over more than 100 years, we divided the data-set into 10 different groups, each one covering 20 years (so that there is overlap between adjacent groups). This allowed us to explore the variation with time of the light-curve, if any. Results are listed in Table~\ref{tab:light_curve_param}. In these computations, we assumed a period of P=65.767~d and T0 (phase of principal maximum) at JD=2413751.75. While most data sets are well fitted by our simple model - supporting the assumption that the binary orbit is indeed circular -, we notice some offsets in the phases for the epochs between 1940 and 1980, with a first excursion to negative values (delays of the light curve with respect to expectation), followed by a similar excursion to positive values; a similar delay is obtained for the REM data set. These offsets however do not imply clear asymmetries in the light curve, with the difference of phase between principal and secondary maximums remaining close to 0.5, The changes in the phase may be as large as 0.08, i.e. $\sim 5$~days, which is much larger than possible uncertainties in the observing epochs and of the variation of the light travel time due to the presence of the M-star companion found by \citet{Mesa2019}\footnote{This last effect is expected to be less than 20~minutes peak-to-valley, and hence is negligible in the present context.}. On the other hand, for the epochs around 2006 (that is, the ASAS and SuperWasp data sets) the light curve does not look symmetric, with the second minimum at a phase later than 0.5. Of course, it is not possible to attribute similar oscillations to changes in the binary orbits. Rather, we are inclined to attribute them to asymmetries in the disk. Variation of the light curve phases with the overall curve still essentially symmetrical, as observed in the epochs between 1940 and 1980, can be attributed to warps in the outer parts of the disk, causing both principal and secondary maximums of the light curve to be delayed/anticipated with respect to the phases where the binary components are in opposition. On the other hand, an asymmetric light curve with phase offsets between principal and secondary maximums different from 0.5 might be attributed to significant changes in the extinction occurring in the inner regions of the disk, close to the binary. 

Inspection of Table~\ref{tab:light_curve_param} also reveals that there is a systematic increase of $h_0$\ with time (see Fig.~\ref{fig:model_par}), that combines with the secular trend for the V magnitude (see Fig-~\ref{fig:AAVSO_sec}). While the effect is not very large, it seems that this secular increase of $h_0$\ is not simply linear, and that there are overimposed shorter term oscillations, with two minimums, one around 1950 and the second one around 2000. This trend might point to a secular increase of the disk thickness, but it may also be explained by a misalignment between the disk and stellar orbital planes and/or by a precession of the disk plane with respect to the orbital plane. The latter might be caused by the M-star companion found by \citet{Mesa2019}, if the orbit of this last object is not co-planar with that of the close binary. While the period of the M-star companion is not well determined, it is likely of the order of 50-100~yrs, and the orbit of the outer part of the disk around the primary is likely shorter than this. The secular variation of the disk extinction supports more a precession scenario rather than a simple misalignment between the disk and the orbital plane of the inner binary, while the shorter term oscillation may indeed be related to the orbital period of the M-star companion.

Recently, \citet{Zhu2019} gave an analytic expression for the precession of a disk due to a massive companion (their Eq. (27)). We re-considered this in the formulation given in eq. (1) of \citep{Garufi2019}. We assumed a mass of 0.25~M$_\odot$ for the companion on a circular orbit with radius of 30 au \citep{Mesa2019}, and that the outer radius of the disk is truncated at the 1:2 resonance with this companion. Furthermore, we assumed that the total mass of the primary (sum of the two components) is 5.34~M$_\odot$\ (see below) and that the inclination between the disk and the companion orbital plane is between 5\textdegree and 10\textdegree (this last parameter is however not critical). With these parameters the precession period is of the order of 1500~yr, and the maximum rate of variation of the disk edge inclination is of a few degree in 100~yr. The interpretation we give of the light curve corresponds to a variation of $h_0$ between 0.23 and 0.26 from 1900 to 2000, that corresponds to a change of $\sim$2\textdegree in the disk inclination, fully compatible with the precession scenario.

On the other hand, the luminosity ratio (in the V-band) between primary and secondary is quite constant at a value of $L_{\rm B}/L_{\rm A}=0.49\pm 0.07$, where the error bar is the rms of values for individual epochs. If we use the relation between mass and $V$\ magnitude for pre-main sequence stars by \citet{Pecaut2013}, the mass ratio between the two components is $0.77\pm 0.05$. The two components should then have a quite similar effective temperature. Neglecting this possible small difference, we may then derive the mass of the two stars that have the same photometric effect of a single star of $3.5\pm 0.5$~M$_\odot$, as considered in Sect-\ref{sec:phot}. We obtain that the two components of the close binary should have masses of $M_A=3.02\pm 0.43$~M$_\odot$ and $M_B=2.32\pm 0.35$~M$_\odot$, respectively. The total mass of the binary should then be $M=5.34\pm 0.8$~M$_\odot$, and given the period, the semimajor axis of the binary should then be 0.56~au.

As noticed by the referee, our photometric model offers the opportunity to test the dependence of the absorption of the disk on wavelength using the simultaneous observations in the $g'r'i'z'$\ obtained with REM. In fact, the best value for the height scale of the disk $h_0$\ is expected to change with wavelength if there is a significant dependence of the extinction on wavelength. We actually found that the same value of $h_0$\ well describes data for all bands considered above, indicating that the extinction is grey. In turn, this suggests that the typical size of the grains within the disk is larger than a few microns. This agrees with the large value of the ratio between total and selective absorption $R_V$\ found in Section~2.

Finally, we notice that according to our model, the disk transmits on average about 45\% of the light from the central binary within the V-band. This implies $A_V \sim 0.9$~mag, i.e., much less than the total absorption toward R~CrA determined in Sect.~\ref{sec:phot} ($A_V \sim 5.47$~mag). On the other hand, the secular variation of the luminosity of R~CrA, with a variation of about 2.5 magnitudes, is much larger. This is tentatively attributed to a larger absorption by precession of the outer portion of the disk. Given the semi-major axis of 27-28 au derived by \citet{Mesa2019}, the period of the M-star companion should then be $61\pm 4$~yr. This agrees well with the short-term oscillation in phase and $h_0$ values shown in Fig.~\ref{fig:model}.

\section{Information from SINFONI spectrum}
\label{sec:spectrum}

\subsection{Emission lines in the spectrum of the M-dwarf companion}

\begin{figure}[htb]
\centering
\includegraphics[width=\columnwidth]{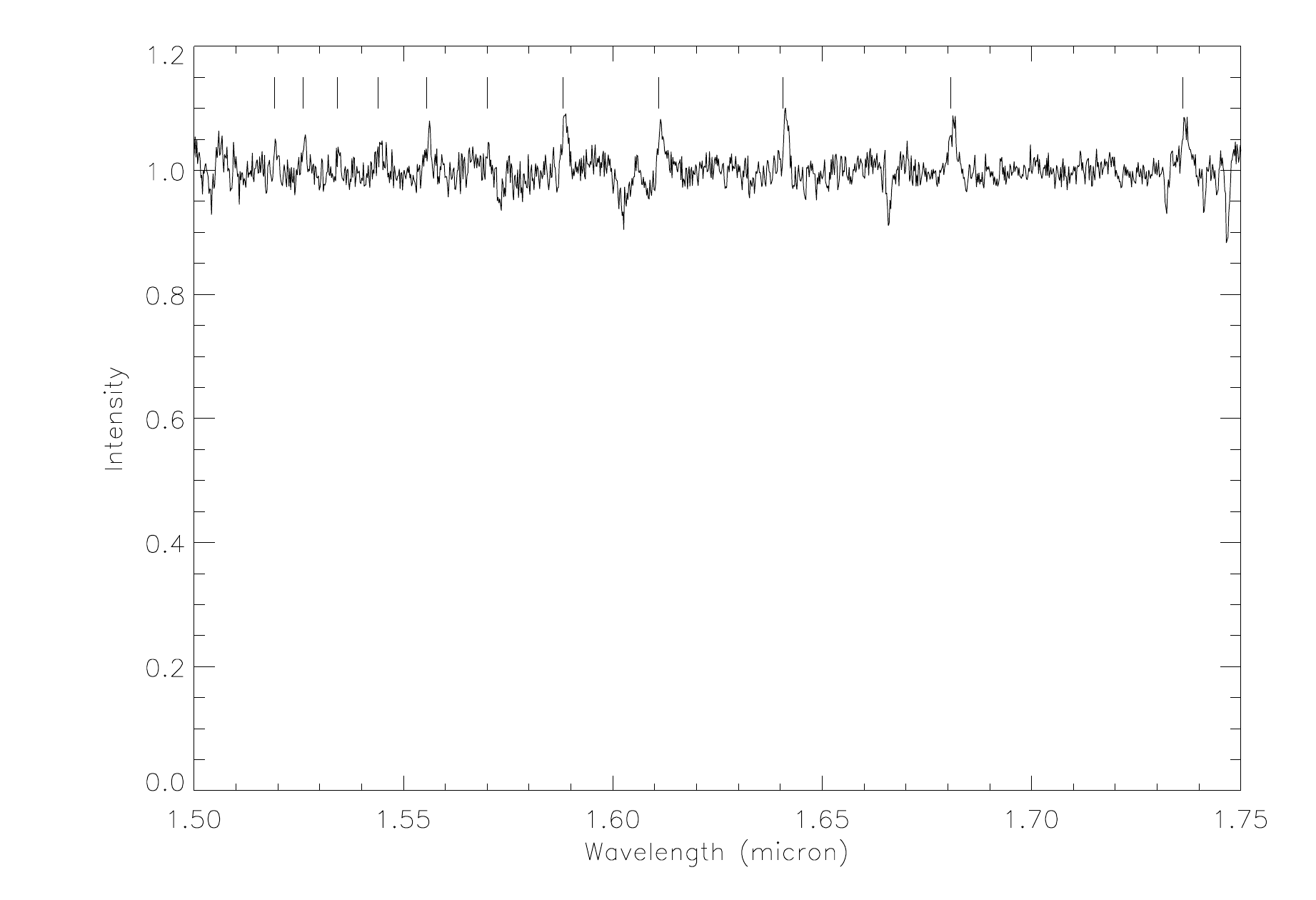}
\includegraphics[width=\columnwidth]{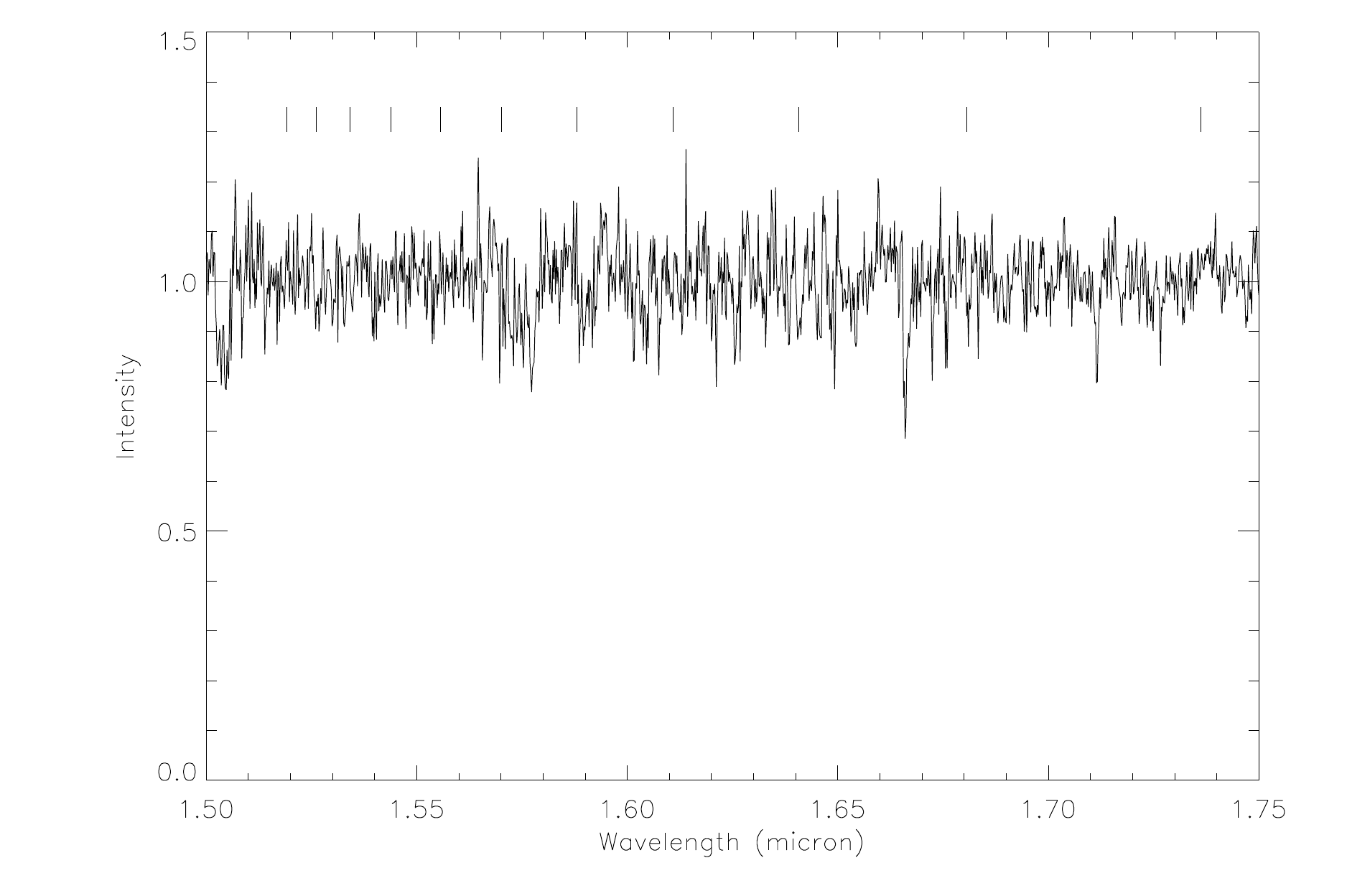}
\caption{H-band spectra extracted from the SINFONI data for the the star (upper panel) and the M-dwarf companion (lower panel). The ticks mark the wavelength corresponding to the H-lines.}
\label{fig:sin_spectra}
\end{figure}

A diffraction limited ($0.8\times 0.8$\arcsec) data set of R~CrA was obtained with the SINFONI integral field spectrograph at the ESO Very Large Telescope \citep{Eisenhauer2003, Bonnet2004} on September 11, 2018 (see \citealp{Mesa2019}). According to our model, this epoch corresponds to phase 0.459, and the spectrum should be dominated by the secondary star. The spectra cover the H-band (from 1.45 to 1.85~$\mu$m); however we only considered the region between 1.5 and 1.75~$\mu$m in our analysis. Figure~\ref{fig:sin_spectra} shows the H-band SINFONI spectra extracted at the locations of the star and of the M-dwarf companion. The wavelengths corresponding of the high-order lines of the Brackett series are also shown. While the star spectrum is dominated by the H-lines, they are not obvious in the spectrum of the companion. We may obtain an upper limit of $\sim 0.2$~nm in the equivalent widths (EWs) of these lines. As a comparison, the EWs of the Paschen$\gamma$\ lines in the spectra of stars of similar mass in Orion is $\sim 0.1$~nm \citep{Rigliaco2012}. Since these lines are expected to be weaker than Paschen$\gamma$, this result is not unexpected.

\subsection{Radial velocities}

%{\bf DA DECIDERE SE LO VOGLIAMO METTERE}

%Per ricavare il punto zero delle velocita' SINFONI, Valentina ha calcolato uno spettro sintetico per le righe telluriche appropriato per i dati SINFONI, e io ho fatto la cross corrrelation di questo spettro con quello osservato per la stella. Non ho usato quello estratto dall immagine cADI, ma quello ottenuto con la sottrazione di un profilo radiale, perche' le righe telluriche sono molto piu' chiare in quest'ultimo. In questo modo si ottiene un offset di $62\pm 4$~km/s nella posizione della stella e $55\pm 4$~km/s nella posizione del jet, nel senso che le righe telluriche appaiono a lunghezza d'onda troppo lunga negli spettri SINFONI. Penso possiamo adottare una correzione media pari a $58\pm 4$~km/s. Avevo ottenuto quanto segue: Se estraggo uno spettro SINFONI nella regione del jet e faccio un fit Gaussiano alla composizione delle righe dell'H, ottengo una velocita' radiale eliocentrica media di $+69\pm 3$~km/s,  con una FWHM=230 km/s. Se faccio lo stesso in prossimita' della stella, la velocita' e' $94\pm 2$~km/s, con una FHWM=172 km/s. Se quindi correggo questi due valori per l'offset trovato per le righe telluriche, ottengo: $V_R=11\pm 5$~km/s in prossimita' del jet $V_R=36\pm 5$~km/s in prossimita' della stella.

%In base al mio  modello spettrofotometrico mi aspetto che all'epoca dello spettro SINFONI, la velocita' della primaria sia Cprossimita' del jet . Le osservazioni sono quindi compatibili con il modello se assumo che le righe dell'H nello spettro SINFONI siano dominate dalla primaria.

The radial velocity of R~CrA is $-36.0\pm 4.9$~km/s \citep{Gontcharov2006}. However, given the complexity of the system, this is unlikely to be the correct velocity of the center of gravity. \citet{Harju1993} gave a systemic velocity of 5.2~km/s for the Coronet cloud with respect to the Local Standard of Rest, meaning an heliocentric value of $-2.1$~km/s. This agrees well with the velocity of $-2$~km/s from the interstellar K~I lines visible in a high resolution \rcra\ spectrum acquired on June 5, 2009, with the FEROS spectrograph that we retrieved from the ESO archive. It also agrees with the value of $-3$~km/s for the mean disk velocity estimated from roto-vibrational CO lines \citep{vanderPlas2015}. This should then be very close to the velocity of the center of gravity of the R~CrA system, that we will assume to be $-2$~km/s.

We measured the radial velocity on the spectrum of the star obtained from the SINFONI data \citep{Mesa2019} using the H lines. We derive a heliocentric velocity $V_R=+8\pm 3$~km/s, that is $V_R=+10\pm 3$~km/s relative to center of gravity of the system. 
%If we repeat the same procedure in the proximity of the jet, we obtain a velocity of $V_R=11\pm 5$~km/s. 
The expected radial velocities with respect to the center of gravity of the system at the epoch of the SINFONI observation in the spectrophotometric model considered above are $V_R=-10$~km/s for the primary and $V_R=+13$~km/s for the secondary. Since the spectrum is expected to be dominated by the secondary. Our result is in agreement with the expected velocity.

Using the SINFONI spectrum we calculated $V_R=+21\pm 7$~km/s for the M-dwarf companion found by \citet{Mesa2019}, that is, $V_R=+23\pm 7$~km/s relative to the center of gravity of the system. This was obtained by cross-correlating the spectrum with that of a dozen template M-dwarfs. This value roughly agrees with the one (about $V_R=+13$~km/s) expected for a clockwise motion around the center of gravity (in agreement with indications from astrometry: see \citealp{Mesa2019}), the near side of the orbit is south-west of the star, and if the total mass of the central binary is 5.34~$M_\odot$, the motion is circular, and the observation was acquired close to quadrature.

\section{Conclusions}
\label{sec:conclusions}

R~CrA, a Herbig AeBe star, is the brightest member of the Coronet association, one of the closest star forming regions at a distance of about 150 pc. High contrast imaging revealed that the star is surrounded by a disk seen at high inclination, has a prominent outflow, and an M-dwarf companion at a separation of about 0.2 arcsec \citep{Mesa2019}. The nature of the central object is still not well assessed. In this paper we provide photometric and spectroscopic evidence that the central object may be a binary with almost equal-mass components. As in many others HAeBe's the stellar emission dominates in the optical and UV, while that from the inner part of the disk dominate in the NIR. The fraction of the total luminosity re-processed by the disk is quite similar among objects of this class, a consequence of the dusty disk truncation by sublimation and of the flaring extension, that does not change too much from object-to-object (see also \citealp{Lazareff2017}). This allows determining the luminosity of the central star from the emission in the NIR, a fact that can be exploited to derive the main properties of the HAeBe stars. We found that the central object in R~CrA is very young (1.5~Myr), quite massive and highly absorbed ($A_V=5.47$). 

We then examined the light curve of R~CrA, for which several data sets are available extending over more than 120 years. This analysis shows a large secular variation - the star is becoming progressively fainter in the last century - and a periodic modulation with a period of $65.767\pm 0.007$ days, in agreement with previous results \citep{Percy2010}. Once phased at this period, the optical light curve shows an almost symmetric variation with a principal and a secondary maximum, while our new data acquired with the REM telescope shows that the luminosity is nearly constant in the NIR (where we essentially see the disk). We interpret this light curve as due to a central binary with two components of $M_A=3.02\pm 0.43$~M$_\odot$ and $M_B=2.32\pm 0.35$~M$_\odot$, whose orbit is seen at an angle that is grazing the circumbinary disk. In analogy with AGB RVb stars \citep{Waelkens1991} and of the T~Tau star KH~15D \citep{Aronow2018}, the variable extinction modulates the light curve. Application of a simple photometric model to the very rich photometric data sets allows us to show that there is some evolution with time in the model parameters pointing toward a progressive increase of the extinction in the last century. We argue that this might be attributed to precession of the outer part of the disk caused by the M-dwarf companion recently discovered with high contrast imaging \citep{Mesa2019}; again, secular luminosity variation related to disk precession makes the case of R~CrA similar to that of KH~15D (see \citealt{Aronow2018} and references therein). R~CrA is then a triple star, with a central quite compact binary composed of two intermediate mass stars and a total mass of $5.34\pm 0.8$~M$_\odot$\ with a period of $65.767\pm 0.007$ days, and a third component of about 0.25~M$_\odot$ and a period of $61\pm 4$~yr.

The \rcra\ SINFONI data described by \citet{Mesa2019} were acquired at a binary phase of 0.459. The H lines seen in emission in this spectrum may be attributed to the low-mass component of the central binary: the measured radial velocity agrees with this interpretation - once an appropriate systemic radial velocity of $-2$~km/s is assumed. The same data set also provides a radial velocity for the M-dwarf companion that agrees with the preliminary orbital solution obtained in  \citet{Mesa2019}.

The present model for R~CrA may be further validated by determining radial velocities and studying line profiles from a sequence of moderate to high spectral resolution spectra taken along the binary orbit; the expected radial velocity variations are of the order of $\pm 70$~km/s. We notice that the spectra of both components should be visible in quadrature, when maximum excursion of radial velocities around the mean values are expected; since lines are quite broad, lines from the two components will be blended and adequate modeling will be required to disentangle the contribution by the two components. The variation of the relative intensity of the emission lines due to the two components will be a further test of the model. In addition, shifts of the photocenter of a few mas are expected due to orbital motion: this should be detectable from GAIA observations. This might perhaps be related to the large discrepancy between the parallax given by GAIA DR2 for R~CrA and the values obtained for all other members of the Coronet clusters. A detailed study of positions obtained by GAIA at individual epochs would then be welcomed, as soon as this data will be available.

%\textcolor{red}{Add conclusions from the FEROS spectrum}

\begin{acknowledgements}
We thank the referee (William Herbst) for useful suggestions that improved the paper. E.S., R.G., D.M., S.D. and V.D. acknowledge support from the ``Progetti Premiali" funding scheme of the Italian Ministry of Education, University, and Research. This work has been supported by the project PRIN-INAF 2016 The Cradle of Life - GENESIS-SKA (General Conditions in Early Planetary Systems for the rise of life with SKA). E.R. is supported by the European Union's Horizon 2020 research and innovation programme under the Marie Sk\l odowska-Curie grant agreement No 664931. We acknowledge with thanks the variable star observations from the AAVSO International Database contributed by observers worldwide and used in this research. This paper makes use of data from the first public release of the WASP data \citep{Butters2010} as provided by the WASP consortium and services at the NASA Exoplanet Archive, which is operated by the California Institute of Technology, under contract with the National Aeronautics and Space Administration under the Exoplanet Exploration Program.
\end{acknowledgements}

%\bibliographystyle{aa}

% Please note that we have included the references to the file aa.dem in order to compile it, but we ask you to:

\begin{thebibliography}{64}
\expandafter\ifx\csname natexlab\endcsname\relax\def\natexlab#1{#1}\fi

\bibitem[{{Alcal{\'a}} {et~al.}(2017){Alcal{\'a}}, {Manara}, {Natta}, {Frasca},
  {Testi}, {Nisini}, {Stelzer}, {Williams}, {Antoniucci}, {Biazzo}, {Covino},
  {Esposito}, {Getman}, \& {Rigliaco}}]{Alcala2017}
{Alcal{\'a}}, J.~M., {Manara}, C.~F., {Natta}, A., {et~al.} 2017, \aap, 600,
  A20

\bibitem[{{Alcal{\'a}} {et~al.}(2008){Alcal{\'a}}, {Spezzi}, {Chapman},
  {Evans}, {Huard}, {J{\o}rgensen}, {Mer{\'{\i}}n}, {Stapelfeldt}, {Covino},
  {Frasca}, {Gandolfi}, \& {Oliveira}}]{Alcala2008}
{Alcal{\'a}}, J.~M., {Spezzi}, L., {Chapman}, N., {et~al.} 2008, \apj, 676, 427

\bibitem[{{Aronow} {et~al.}(2018){Aronow}, {Herbst}, {Hughes}, {Wilner}, \&
  {Winn}}]{Aronow2018}
{Aronow}, R.~A., {Herbst}, W., {Hughes}, A.~M., {Wilner}, D.~J., \& {Winn},
  J.~N. 2018, \aj, 155, 47

\bibitem[{{Bellingham} \& {Rossano}(1980)}]{Bellingham1980}
{Bellingham}, J.~G. \& {Rossano}, G.~S. 1980, \aj, 85, 555

\bibitem[{{Bessell} \& {Brett}(1988)}]{Bessell1988}
{Bessell}, M.~S. \& {Brett}, J.~M. 1988, \pasp, 100, 1134

\bibitem[{{Beuzit} {et~al.}(2019){Beuzit}, {Vigan}, {Mouillet}, {Dohlen},
  {Gratton}, {Boccaletti}, {Sauvage}, {Schmid}, {Langlois}, {Petit},
  {Baruffolo}, {Feldt}, {Milli}, {Wahhaj}, {Abe}, {Anselmi}, {Antichi},
  {Barette}, {Baudrand}, {Baudoz}, {Bazzon}, {Bernardi}, {Blanchard}, {Brast},
  {Bruno}, {Buey}, {Carbillet}, {Carle}, {Cascone}, {Chapron}, {Chauvin},
  {Charton}, {Claudi}, {Costille}, {De Caprio}, {Delboulb{\'e}}, {Desidera},
  {Dominik}, {Downing}, {Dupuis}, {Fabron}, {Fantinel}, {Farisato},
  {Feautrier}, {Fedrigo}, {Fusco}, {Gigan}, {Ginski}, {Girard}, {Giro},
  {Gisler}, {Gluck}, {Gry}, {Henning}, {Hubin}, {Hugot}, {Incorvaia}, {Jaquet},
  {Kasper}, {Lagadec}, {Lagrange}, {Le Coroller}, {Le Mignant}, {Le Ruyet},
  {Lessio}, {Lizon}, {Llored}, {Lundin}, {Madec}, {Magnard}, {Marteaud},
  {Martinez}, {Maurel}, {M{\'e}nard}, {Mesa}, {M{\"o}ller-Nilsson}, {Moulin},
  {Moutou}, {Orign{\'e}}, {Parisot}, {Pavlov}, {Perret}, {Pragt}, {Puget},
  {Rabou}, {Ramos}, {Reess}, {Rigal}, {Rochat}, {Roelfsema}, {Rousset}, {Roux},
  {Saisse}, {Salasnich}, {Santambrogio}, {Scuderi}, {Segransan}, {Sevin},
  {Siebenmorgen}, {Soenke}, {Stadler}, {Suarez}, {Tiph{\`e}ne}, {Turatto},
  {Udry}, {Vakili}, {Waters}, {Weber}, {Wildi}, {Zins}, \&
  {Zurlo}}]{Beuzit2019}
{Beuzit}, J.-L., {Vigan}, A., {Mouillet}, D., {et~al.} 2019, arXiv e-prints
  [\eprint[arXiv]{1902.04080}]

\bibitem[{{Bibo} {et~al.}(1992){Bibo}, {The}, \& {Dawanas}}]{Bibo1992}
{Bibo}, E.~A., {The}, P.~S., \& {Dawanas}, D.~N. 1992, \aap, 260, 293

\bibitem[{{Blondel} \& {Djie}(2006)}]{Blondel2006}
{Blondel}, P.~F.~C. \& {Djie}, H.~R.~E.~T.~A. 2006, \aap, 456, 1045

\bibitem[{{Bonnet} {et~al.}(2004){Bonnet}, {Abuter}, {Baker}, {Bornemann},
  {Brown}, {Castillo}, {Conzelmann}, {Damster}, {Davies}, {Delabre},
  {Donaldson}, {Dumas}, {Eisenhauer}, {Elswijk}, {Fedrigo}, {Finger},
  {Gemperlein}, {Genzel}, {Gilbert}, {Gillet}, {Goldbrunner}, {Horrobin}, {Ter
  Horst}, {Huber}, {Hubin}, {Iserlohe}, {Kaufer}, {Kissler-Patig}, {Kragt},
  {Kroes}, {Lehnert}, {Lieb}, {Liske}, {Lizon}, {Lutz}, {Modigliani}, {Monnet},
  {Nesvadba}, {Patig}, {Pragt}, {Reunanen}, {R{\"o}hrle}, {Rossi}, {Schmutzer},
  {Schoenmaker}, {Schreiber}, {Stroebele}, {Szeifert}, {Tacconi}, {Tecza},
  {Thatte}, {Tordo}, {van der Werf}, \& {Weisz}}]{Bonnet2004}
{Bonnet}, H., {Abuter}, R., {Baker}, A., {et~al.} 2004, The Messenger, 117, 17

\bibitem[{{Butters} {et~al.}(2010){Butters}, {West}, {Anderson}, {Collier
  Cameron}, {Clarkson}, {Enoch}, {Haswell}, {Hellier}, {Horne}, {Joshi},
  {Kane}, {Lister}, {Maxted}, {Parley}, {Pollacco}, {Smalley}, {Street},
  {Todd}, {Wheatley}, \& {Wilson}}]{Butters2010}
{Butters}, O.~W., {West}, R.~G., {Anderson}, D.~R., {et~al.} 2010, \aap, 520,
  L10

\bibitem[{{Cardelli} {et~al.}(1989){Cardelli}, {Clayton}, \&
  {Mathis}}]{Cardelli1989}
{Cardelli}, J.~A., {Clayton}, G.~C., \& {Mathis}, J.~S. 1989, \apj, 345, 245

\bibitem[{{Chauvin} {et~al.}(2017){Chauvin}, {Desidera}, {Lagrange}, {Vigan},
  {Feldt}, {Gratton}, {Langlois}, {Cheetham}, {Bonnefoy}, \&
  {Meyer}}]{Chauvin2017}
{Chauvin}, G., {Desidera}, S., {Lagrange}, A.-M., {et~al.} 2017, in SF2A-2017:
  Proceedings of the Annual meeting of the French Society of Astronomy and
  Astrophysics, ed. C.~{Reyl{\'e}}, P.~{Di Matteo}, F.~{Herpin}, E.~{Lagadec},
  A.~{Lan{\c c}on}, Z.~{Meliani}, \& F.~{Royer}, 331--335

\bibitem[{{Chen} {et~al.}(2012){Chen}, {Pecaut}, {Mamajek}, {Su}, \&
  {Bitner}}]{Chen2012}
{Chen}, C.~H., {Pecaut}, M., {Mamajek}, E.~E., {Su}, K.~Y.~L., \& {Bitner}, M.
  2012, \apj, 756, 133

\bibitem[{{Chen} {et~al.}(1997){Chen}, {Grenfell}, {Myers}, {P.~C.}, \&
  {Hughes}}]{Chen1997}
{Chen}, H., {Grenfell}, T.~G., {Myers}, {P.~C.}, \& {Hughes}, J.~D. 1997, \apj,
  478, 295

\bibitem[{{Chincarini} {et~al.}(2003){Chincarini}, {Zerbi}, {Antonelli},
  {Conconi}, {Cutispoto}, {Covino}, {D'Alessio}, {de Ugarte Postigo},
  {Molinari}, {Nicastro}, {Tosti}, {Vitali}, {Mazzoleni}, {Sciuto}, {Stefanon},
  {Jordan}, {Burderi}, {Campana}, {Danziger}, {di Paola}, {Fernandez-Soto},
  {Fiore}, {Ghisellini}, {Goldoni}, {Israel}, {Lorenzetti}, {McBreen},
  {Masetti}, {Messina}, {Meurs}, {Monfardini}, {Palazzi}, {Paul}, {Pian},
  {Rodono}, {Stella}, {Tagliaferri}, {Testa}, \& {Vergani}}]{Chincarini2003}
{Chincarini}, G., {Zerbi}, F., {Antonelli}, A., {et~al.} 2003, The Messenger,
  113, 40

\bibitem[{{Cutri} {et~al.}(2003){Cutri}, {Skrutskie}, {van Dyk}, {Beichman},
  {Carpenter}, {Chester}, {Cambresy}, {Evans}, {Fowler}, {Gizis}, {Howard},
  {Huchra}, {Jarrett}, {Kopan}, {Kirkpatrick}, {Light}, {Marsh}, {McCallon},
  {Schneider}, {Stiening}, {Sykes}, {Weinberg}, {Wheaton}, {Wheelock}, \&
  {Zacarias}}]{Cutri2003}
{Cutri}, R.~M., {Skrutskie}, M.~F., {van Dyk}, S., {et~al.} 2003, VizieR Online
  Data Catalog, 2246

\bibitem[{{Cutri}(2013)}]{Cutri2013}
{Cutri}, R.~M.~e. 2013, VizieR Online Data Catalog, 2328

\bibitem[{{Ducati}(2002)}]{Ducati2002}
{Ducati}, J.~R. 2002, VizieR Online Data Catalog, 2237

\bibitem[{{Eiroa} {et~al.}(2002){Eiroa}, {Oudmaijer}, {Davies}, {de Winter},
  {Garz{\'o}n}, {Palacios}, {Alberdi}, {Ferlet}, {Grady}, {Collier Cameron},
  {Deeg}, {Harris}, {Horne}, {Mer{\'\i}n}, {Miranda}, {Montesinos}, {Mora},
  {Penny}, {Quirrenbach}, {Rauer}, {Schneider}, {Solano}, {Tsapras}, \&
  {Wesselius}}]{Eiroa2002}
{Eiroa}, C., {Oudmaijer}, R.~D., {Davies}, J.~K., {et~al.} 2002, \aap, 384,
  1038

\bibitem[{{Eisenhauer} {et~al.}(2003){Eisenhauer}, {Abuter}, {Bickert},
  {Biancat-Marchet}, {Bonnet}, {Brynnel}, {Conzelmann}, {Delabre}, {Donaldson},
  {Farinato}, {Fedrigo}, {Genzel}, {Hubin}, {Iserlohe}, {Kasper},
  {Kissler-Patig}, {Monnet}, {Roehrle}, {Schreiber}, {Stroebele}, {Tecza},
  {Thatte}, \& {Weisz}}]{Eisenhauer2003}
{Eisenhauer}, F., {Abuter}, R., {Bickert}, K., {et~al.} 2003, in \procspie,
  Vol. 4841, Instrument Design and Performance for Optical/Infrared
  Ground-based Telescopes, ed. M.~{Iye} \& A.~F.~M. {Moorwood}, 1548--1561

\bibitem[{{Fairlamb} {et~al.}(2015){Fairlamb}, {Oudmaijer},
  {Mendigut{\'{\i}}a}, {Ilee}, \& {van den Ancker}}]{Fairlamb2015}
{Fairlamb}, J.~R., {Oudmaijer}, R.~D., {Mendigut{\'{\i}}a}, I., {Ilee}, J.~D.,
  \& {van den Ancker}, M.~E. 2015, \mnras, 453, 976

\bibitem[{{Forbrich} {et~al.}(2006){Forbrich}, {Preibisch}, \&
  {Menten}}]{Forbrich2006}
{Forbrich}, J., {Preibisch}, T., \& {Menten}, K.~M. 2006, \aap, 446, 155

\bibitem[{{Gaia Collaboration} {et~al.}(2018){Gaia Collaboration}, {Brown},
  {Vallenari}, {Prusti}, {de Bruijne}, {Babusiaux}, {Bailer-Jones}, {Biermann},
  {Evans}, {Eyer}, \& et~al.}]{Gaia2018}
{Gaia Collaboration}, {Brown}, A.~G.~A., {Vallenari}, A., {et~al.} 2018, \aap,
  616, A1

\bibitem[{{Garufi} {et~al.}(2019){Garufi}, {Podio}, {Bacciotti}, {Antoniucci},
  {Boccaletti}, {Codella}, {Dougados}, {Menard}, {Mesa}, {Meyer}, {Nisini},
  {Schmid}, {Stolker}, {Baudino}, {Biller}, {Bonavita}, {Bonnefoy},
  {Cantalloube}, {Chauvin}, {Cheetham}, {Desidera}, {D'Orazi}, {Feldt},
  {Galicher}, {Grandjean}, {Gratton}, {Hagelberg}, {Lagrange}, {Langlois},
  {Lannier}, {Lazzoni}, {Maire}, {Perrot}, {Rickman}, {Schmidt}, {Vigan},
  {Zurlo}, {Delboulbe}, {Le Mignant}, {Fantinel}, {Moeller-Nilsson}, {Weber},
  \& {Sauvage}}]{Garufi2019}
{Garufi}, A., {Podio}, L., {Bacciotti}, F., {et~al.} 2019, arXiv e-prints,
  arXiv:1906.06910

\bibitem[{{Gontcharov}(2006)}]{Gontcharov2006}
{Gontcharov}, G.~A. 2006, Astronomy Letters, 32, 759

\bibitem[{{Gray} {et~al.}(2006){Gray}, {Corbally}, {Garrison}, {McFadden},
  {Bubar}, {McGahee}, {O'Donoghue}, \& {Knox}}]{Gray2006}
{Gray}, R.~O., {Corbally}, C.~J., {Garrison}, R.~F., {et~al.} 2006, \aj, 132,
  161

\bibitem[{{Hamann} \& {Persson}(1992)}]{Hamann1992}
{Hamann}, F. \& {Persson}, S.~E. 1992, \apjs, 82, 285

\bibitem[{{Harju} {et~al.}(1993){Harju}, {Haikala}, {Mattila}, {Mauersberger},
  {Booth}, \& {Nordh}}]{Harju1993}
{Harju}, J., {Haikala}, L.~K., {Mattila}, K., {et~al.} 1993, \aap, 278, 569

\bibitem[{{Herbig}(1960)}]{Herbig1960}
{Herbig}, G.~H. 1960, \apjs, 4, 337

\bibitem[{{Herbst} \& {Shevchenko}(1999)}]{Herbst1999A}
{Herbst}, W. \& {Shevchenko}, V.~S. 1999, \aj, 118, 1043

\bibitem[{{Hern{\'a}ndez} {et~al.}(2004){Hern{\'a}ndez}, {Calvet},
  {Brice{\~n}o}, {Hartmann}, \& {Berlind}}]{Hernandez2004}
{Hern{\'a}ndez}, J., {Calvet}, N., {Brice{\~n}o}, C., {Hartmann}, L., \&
  {Berlind}, P. 2004, \aj, 127, 1682

\bibitem[{{Hillenbrand} {et~al.}(1992){Hillenbrand}, {Strom}, {Vrba}, \&
  {Keene}}]{Hillenbrand1992}
{Hillenbrand}, L.~A., {Strom}, S.~E., {Vrba}, F.~J., \& {Keene}, J. 1992, \apj,
  397, 613

\bibitem[{{Johnson} {et~al.}(2010){Johnson}, {Aller}, {Howard}, \&
  {Crepp}}]{Johnson2010}
{Johnson}, J.~A., {Aller}, K.~M., {Howard}, A.~W., \& {Crepp}, J.~R. 2010,
  \pasp, 122, 905

\bibitem[{{Kenyon} \& {Hartmann}(1987)}]{Kenyon1987}
{Kenyon}, S.~J. \& {Hartmann}, L. 1987, \apj, 323, 714

\bibitem[{{Kochanek} {et~al.}(2017){Kochanek}, {Shappee}, {Stanek}, {Holoien},
  {Thompson}, {Prieto}, {Dong}, {Shields}, {Will}, {Britt}, {Perzanowski}, \&
  {Pojma{\'n}ski}}]{Kochanek2017}
{Kochanek}, C.~S., {Shappee}, B.~J., {Stanek}, K.~Z., {et~al.} 2017, \pasp,
  129, 104502

\bibitem[{{Koen} {et~al.}(2010){Koen}, {Kilkenny}, {van Wyk}, \&
  {Marang}}]{Koen2010}
{Koen}, C., {Kilkenny}, D., {van Wyk}, F., \& {Marang}, F. 2010, \mnras, 403,
  1949

\bibitem[{{Kraus} {et~al.}(2009){Kraus}, {Hofmann}, {Malbet}, {Meilland},
  {Natta}, {Schertl}, {Stee}, \& {Weigelt}}]{Kraus2009}
{Kraus}, S., {Hofmann}, K.-H., {Malbet}, F., {et~al.} 2009, \aap, 508, 787

\bibitem[{{Lazareff} {et~al.}(2017){Lazareff}, {Berger}, {Kluska}, {Le
  Bouquin}, {Benisty}, {Malbet}, {Koen}, {Pinte}, {Thi}, {Absil}, {Baron},
  {Delboulb{\'e}}, {Duvert}, {Isella}, {Jocou}, {Juhasz}, {Kraus}, {Lachaume},
  {M{\'e}nard}, {Millan-Gabet}, {Monnier}, {Moulin}, {Perraut}, {Rochat},
  {Soulez}, {Tallon}, {Thi{\'e}baut}, {Traub}, \& {Zins}}]{Lazareff2017}
{Lazareff}, B., {Berger}, J.-P., {Kluska}, J., {et~al.} 2017, \aap, 599, A85

\bibitem[{{Manoj} \& {Bhatt}(2005)}]{Manoj2005b}
{Manoj}, P. \& {Bhatt}, H.~C. 2005, \aap, 429, 525

\bibitem[{{Marshall} {et~al.}(2014){Marshall}, {Moro-Mart{\'{\i}}n}, {Eiroa},
  {Kennedy}, {Mora}, {Sibthorpe}, {Lestrade}, {Maldonado}, {Sanz-Forcada},
  {Wyatt}, {Matthews}, {Horner}, {Montesinos}, {Bryden}, {del Burgo},
  {Greaves}, {Ivison}, {Meeus}, {Olofsson}, {Pilbratt}, \&
  {White}}]{Marshall2014}
{Marshall}, J.~P., {Moro-Mart{\'{\i}}n}, A., {Eiroa}, C., {et~al.} 2014, \aap,
  565, A15

\bibitem[{{Meijer} {et~al.}(2008){Meijer}, {Dominik}, {de Koter}, {Dullemond},
  {van Boekel}, \& {Waters}}]{Meijer2008}
{Meijer}, J., {Dominik}, C., {de Koter}, A., {et~al.} 2008, \aap, 492, 451

\bibitem[{{Mer{\'{\i}}n} {et~al.}(2008){Mer{\'{\i}}n}, {J{\o}rgensen},
  {Spezzi}, {Alcal{\'a}}, {Evans}, {Harvey}, {Prusti}, {Chapman}, {Huard}, {van
  Dishoeck}, \& {Comer{\'o}n}}]{Merin2008}
{Mer{\'{\i}}n}, B., {J{\o}rgensen}, J., {Spezzi}, L., {et~al.} 2008, \apjs,
  177, 551

\bibitem[{{Mer{\'{\i}}n} {et~al.}(2004){Mer{\'{\i}}n}, {Montesinos}, {Eiroa},
  {Solano}, {Mora}, {D'Alessio}, {Calvet}, {Oudmaijer}, {de Winter}, {Davies},
  {Harris}, {Collier Cameron}, {Deeg}, {Ferlet}, {Garz{\'o}n}, {Grady},
  {Horne}, {Miranda}, {Palacios}, {Penny}, {Quirrenbach}, {Rauer}, {Schneider},
  \& {Wesselius}}]{Merin2004}
{Mer{\'{\i}}n}, B., {Montesinos}, B., {Eiroa}, C., {et~al.} 2004, \aap, 419,
  301

\bibitem[{{Mesa} {et~al.}(2019){Mesa}, {Bonnefoy}, {Gratton}, {Van Der Plas},
  {D'Orazi}, {Sissa}, {Zurlo}, {Rigliaco}, {Schmidt}, {Langlois}, {Vigan},
  {Ubeira Gabellini}, {Desidera}, {Antoniucci}, {Barbieri}, {Benisty},
  {Boccaletti}, {Claudi}, {Fedele}, {Gasparri}, {Henning}, {Kasper},
  {Lagrange}, {Lazzoni}, {Lodato}, {Maire}, {Manara}, {Meyer}, {Reggiani},
  {Samland}, {Van den Ancker}, {Chauvin}, {Cheetham}, {Feldt}, {Hugot},
  {Janson}, {Ligi}, {M{\"o}ller-Nilsson}, {Petit}, {Rickman}, {Rigal}, \&
  {Wildi}}]{Mesa2019}
{Mesa}, D., {Bonnefoy}, M., {Gratton}, R., {et~al.} 2019, \aap, 624, A4

\bibitem[{{Montesinos} {et~al.}(2009){Montesinos}, {Eiroa}, {Mora}, \&
  {Mer{\'{\i}}n}}]{Montesinos2009}
{Montesinos}, B., {Eiroa}, C., {Mora}, A., \& {Mer{\'{\i}}n}, B. 2009, \aap,
  495, 901

\bibitem[{{Neuh{\"a}user} \& {Forbrich}(2008)}]{Neuhauser2008}
{Neuh{\"a}user}, R. \& {Forbrich}, J. 2008, in Handbook of Star Forming
  Regions, Volume II, ed. B.~{Reipurth}, 735

\bibitem[{{Nielsen} {et~al.}(2019){Nielsen}, {De Rosa}, {Macintosh}, {Wang},
  {Ruffio}, {Chiang}, {Marley}, {Saumon}, {Savransky}, {Ammons}, {Bailey},
  {Barman}, {Blain}, {Bulger}, {Chilcote}, {Cotten}, {Czekala}, {Doyon},
  {Duchene}, {Esposito}, {Fabrycky}, {Fitzgerald}, {Follette}, {Fortney},
  {Gerard}, {Goodsell}, {Graham}, {Greenbaum}, {Hibon}, {Hinkley}, {Hirsch},
  {Hom}, {Hung}, {Dawson}, {Ingraham}, {Kalas}, {Konopacky}, {Larkin}, {Lee},
  {Lin}, {Maire}, {Marchis}, {Marois}, {Metchev}, {Millar-Blanchaer},
  {Morzinski}, {Oppenheimer}, {Palmer}, {Patience}, {Perrin}, {Poyneer},
  {Pueyo}, {Rafikov}, {Rajan}, {Rameau}, {Rantakyro}, {Ren}, {Schneider},
  {Sivaramakrishnan}, {Song}, {Soummer}, {Tallis}, {Thomas}, {Ward-Duong}, \&
  {Wolff}}]{Nielsen2019}
{Nielsen}, E.~L., {De Rosa}, R.~J., {Macintosh}, B., {et~al.} 2019, arXiv
  e-prints [\eprint[arXiv]{1904.05358}]

\bibitem[{{Oudmaijer} {et~al.}(2001){Oudmaijer}, {Palacios}, {Eiroa}, {Davies},
  {de Winter}, {Ferlet}, {Garz{\'o}n}, {Grady}, {Collier Cameron}, {Deeg},
  {Harris}, {Horne}, {Mer{\'\i}n}, {Miranda}, {Montesinos}, {Mora}, {Penny},
  {Quirrenbach}, {Rauer}, {Schneider}, {Solano}, {Tsapras}, \&
  {Wesselius}}]{Oudmaijer2001}
{Oudmaijer}, R.~D., {Palacios}, J., {Eiroa}, C., {et~al.} 2001, \aap, 379, 564

\bibitem[{{Pecaut} \& {Mamajek}(2013)}]{Pecaut2013}
{Pecaut}, M.~J. \& {Mamajek}, E.~E. 2013, \apjs, 208, 9

\bibitem[{{Percy} {et~al.}(2010){Percy}, {Grynko}, {Seneviratne}, \&
  {Herbst}}]{Percy2010}
{Percy}, J.~R., {Grynko}, S., {Seneviratne}, R., \& {Herbst}, W. 2010, \pasp,
  122, 753

\bibitem[{{Perez} \& {Grady}(1997)}]{Perez1997}
{Perez}, M.~R. \& {Grady}, C.~A. 1997, \ssr, 82, 407

\bibitem[{{Pojmanski}(1997)}]{Pojmanski1997}
{Pojmanski}, G. 1997, \actaa, 47, 467

\bibitem[{{Rigliaco} {et~al.}(2012){Rigliaco}, {Natta}, {Testi}, {Randich},
  {Alcal{\`a}}, {Covino}, \& {Stelzer}}]{Rigliaco2012}
{Rigliaco}, E., {Natta}, A., {Testi}, L., {et~al.} 2012, \aap, 548, A56

\bibitem[{{Shappee} {et~al.}(2014){Shappee}, {Prieto}, {Grupe}, {Kochanek},
  {Stanek}, {De Rosa}, {Mathur}, {Zu}, {Peterson}, {Pogge}, {Komossa}, {Im},
  {Jencson}, {Holoien}, {Basu}, {Beacom}, {Szczygie{\l}}, {Brimacombe},
  {Adams}, {Campillay}, {Choi}, {Contreras}, {Dietrich}, {Dubberley},
  {Elphick}, {Foale}, {Giustini}, {Gonzalez}, {Hawkins}, {Howell}, {Hsiao},
  {Koss}, {Leighly}, {Morrell}, {Mudd}, {Mullins}, {Nugent}, {Parrent},
  {Phillips}, {Pojmanski}, {Rosing}, {Ross}, {Sand}, {Terndrup}, {Valenti},
  {Walker}, \& {Yoon}}]{Shappee2014}
{Shappee}, B.~J., {Prieto}, J.~L., {Grupe}, D., {et~al.} 2014, \apj, 788, 48

\bibitem[{{Skrutskie} {et~al.}(2006){Skrutskie}, {Cutri}, {Stiening},
  {Weinberg}, {Schneider}, {Carpenter}, {Beichman}, {Capps}, {Chester},
  {Elias}, {Huchra}, {Liebert}, {Lonsdale}, {Monet}, {Price}, {Seitzer},
  {Jarrett}, {Kirkpatrick}, {Gizis}, {Howard}, {Evans}, {Fowler}, {Fullmer},
  {Hurt}, {Light}, {Kopan}, {Marsh}, {McCallon}, {Tam}, {Van Dyk}, \&
  {Wheelock}}]{Skrutskie2006}
{Skrutskie}, M.~F., {Cutri}, R.~M., {Stiening}, R., {et~al.} 2006, \aj, 131,
  1163

\bibitem[{{Takami} {et~al.}(2003){Takami}, {Bailey}, \&
  {Chrysostomou}}]{Takami2003}
{Takami}, M., {Bailey}, J., \& {Chrysostomou}, A. 2003, \aap, 397, 675

\bibitem[{{The}(1994)}]{The1994}
{The}, P.~S. 1994, in Astronomical Society of the Pacific Conference Series,
  Vol.~62, The Nature and Evolutionary Status of Herbig Ae/Be Stars, ed. P.~S.
  {The}, M.~R. {Perez}, \& E.~P.~J. {van den Heuvel}, 23

\bibitem[{{Tognelli} {et~al.}(2011){Tognelli}, {Prada Moroni}, \&
  {Degl'Innocenti}}]{Tognelli2011}
{Tognelli}, E., {Prada Moroni}, P.~G., \& {Degl'Innocenti}, S. 2011, \aap, 533,
  A109

\bibitem[{{van der Plas} {et~al.}(2015){van der Plas}, {van den Ancker},
  {Waters}, \& {Dominik}}]{vanderPlas2015}
{van der Plas}, G., {van den Ancker}, M.~E., {Waters}, L.~B.~F.~M., \&
  {Dominik}, C. 2015, \aap, 574, A75

\bibitem[{{Vioque} {et~al.}(2018){Vioque}, {Oudmaijer}, {Baines},
  {Mendigut{\'{\i}}a}, \& {P{\'e}rez-Mart{\'{\i}}nez}}]{vioque2018}
{Vioque}, M., {Oudmaijer}, R.~D., {Baines}, D., {Mendigut{\'{\i}}a}, I., \&
  {P{\'e}rez-Mart{\'{\i}}nez}, R. 2018, \aap, 620, A128

\bibitem[{{Waelkens} {et~al.}(1991){Waelkens}, {Lamers}, {Waters}, {Rufener},
  {Trams}, {Le Bertre}, {Ferlet}, \& {Vidal-Madjar}}]{Waelkens1991}
{Waelkens}, C., {Lamers}, H.~J.~G.~L.~M., {Waters}, L.~B.~F.~M., {et~al.} 1991,
  \aap, 242, 433

\bibitem[{{Waelkens} {et~al.}(1996){Waelkens}, {Van Winckel}, {Waters}, \&
  {Bakker}}]{Waelkens1996}
{Waelkens}, C., {Van Winckel}, H., {Waters}, L.~B.~F.~M., \& {Bakker}, E.~J.
  1996, \aap, 314, L17

\bibitem[{{Zechmeister} \& {K{\"u}rster}(2009)}]{Zechmeister2009}
{Zechmeister}, M. \& {K{\"u}rster}, M. 2009, \aap, 496, 577

\bibitem[{{Zhu}(2019)}]{Zhu2019}
{Zhu}, Z. 2019, \mnras, 483, 4221

\end{thebibliography}

%\begin{appendix}
\section{Appendix. Model fit to the photometric curves}

\begin{figure*}[htb]
    \centering
    \begin{tabular}{cc}
    \includegraphics[width=0.8\columnwidth]{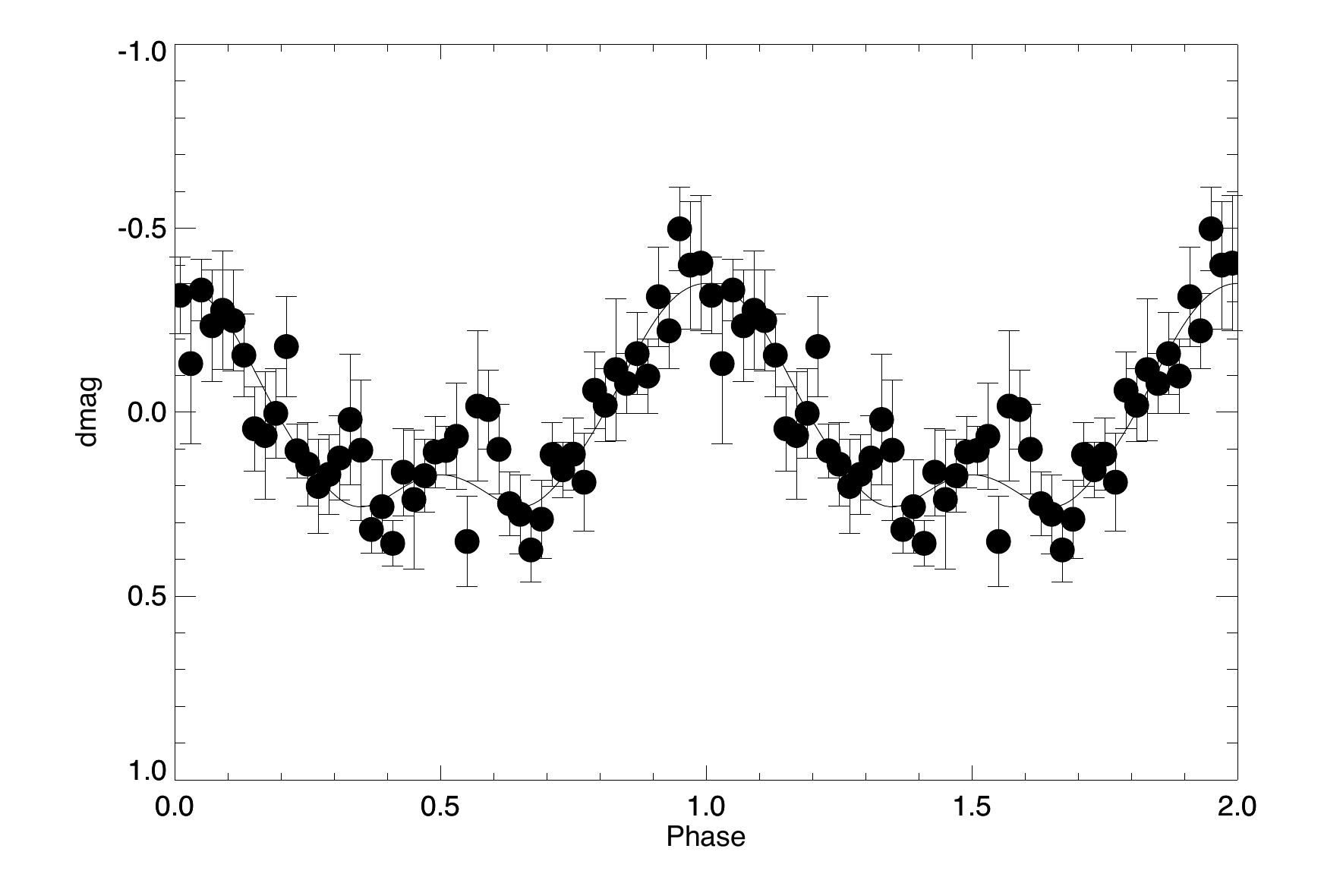}&
    \includegraphics[width=0.8\columnwidth]{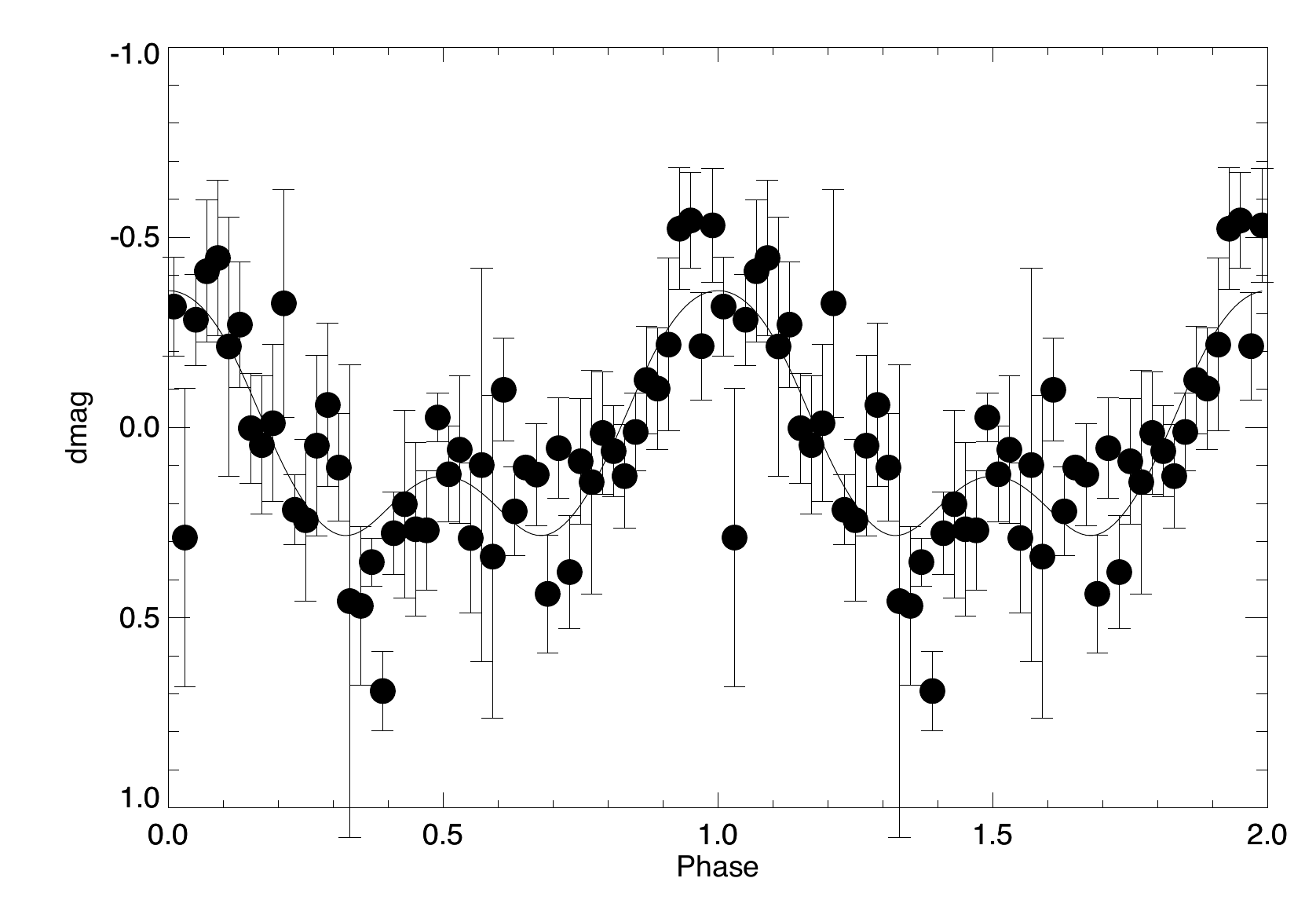}\\
    \includegraphics[width=0.8\columnwidth]{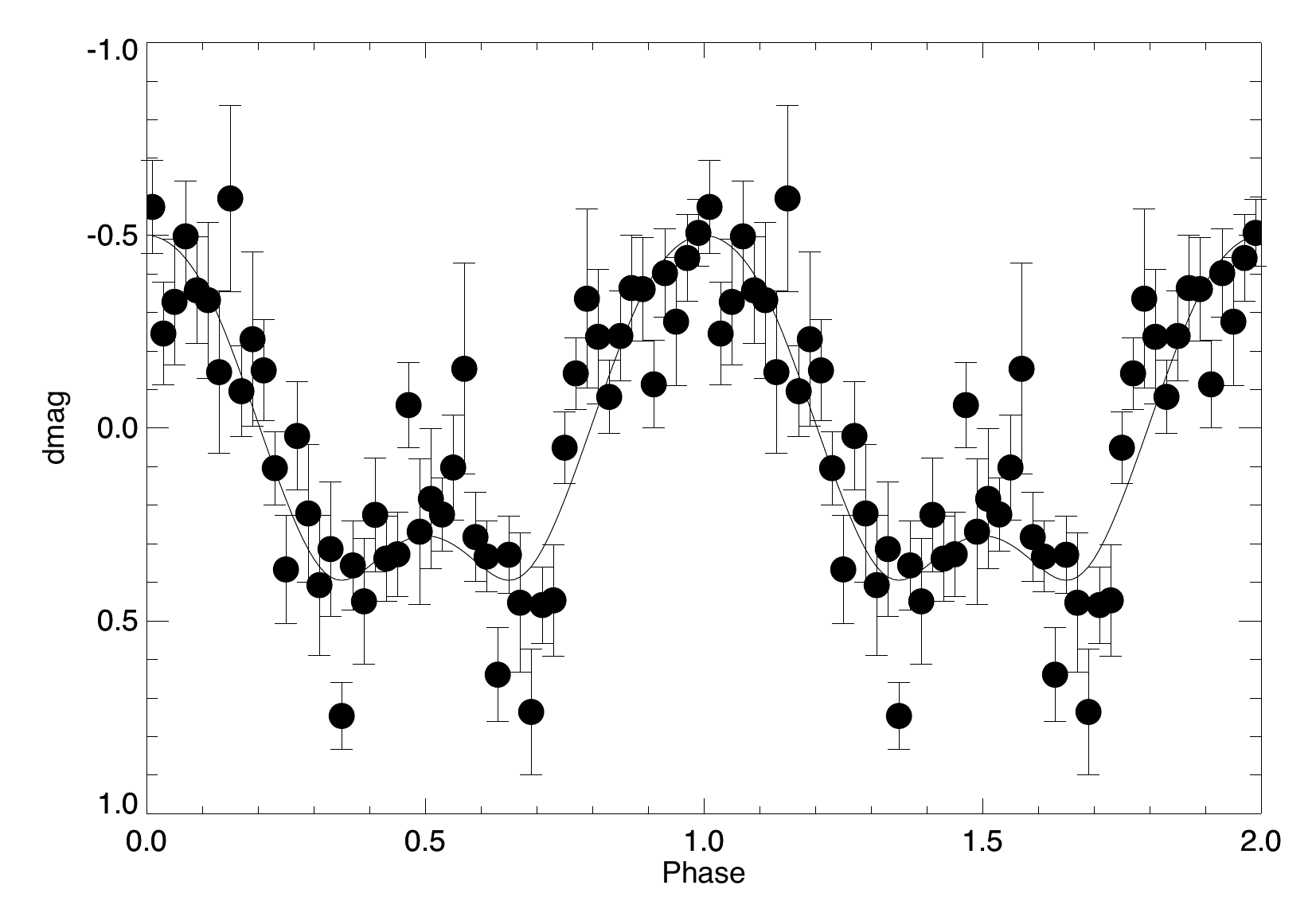}&
    \includegraphics[width=0.8\columnwidth]{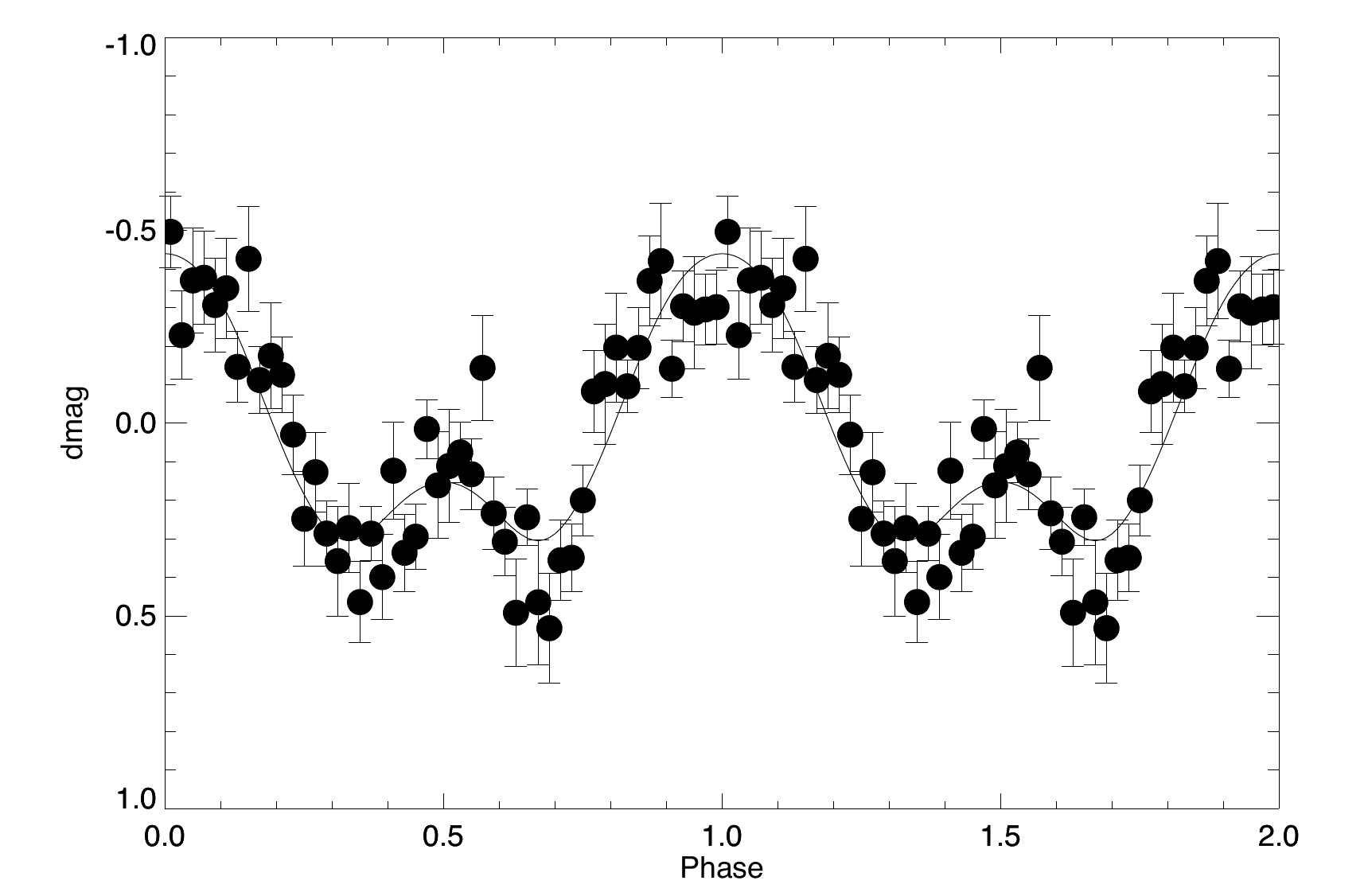}\\
    \includegraphics[width=0.8\columnwidth]{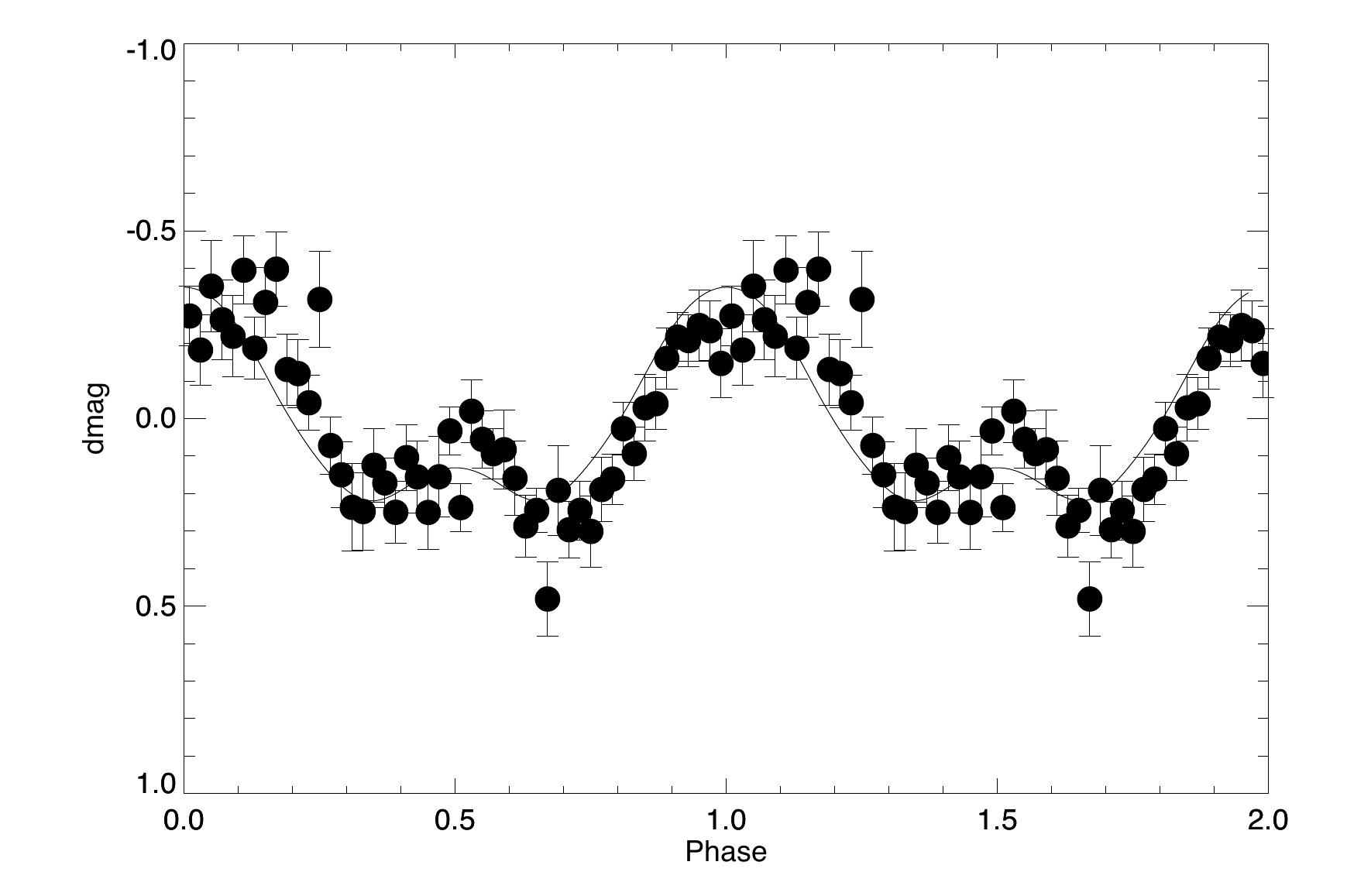}&
    \includegraphics[width=0.8\columnwidth]{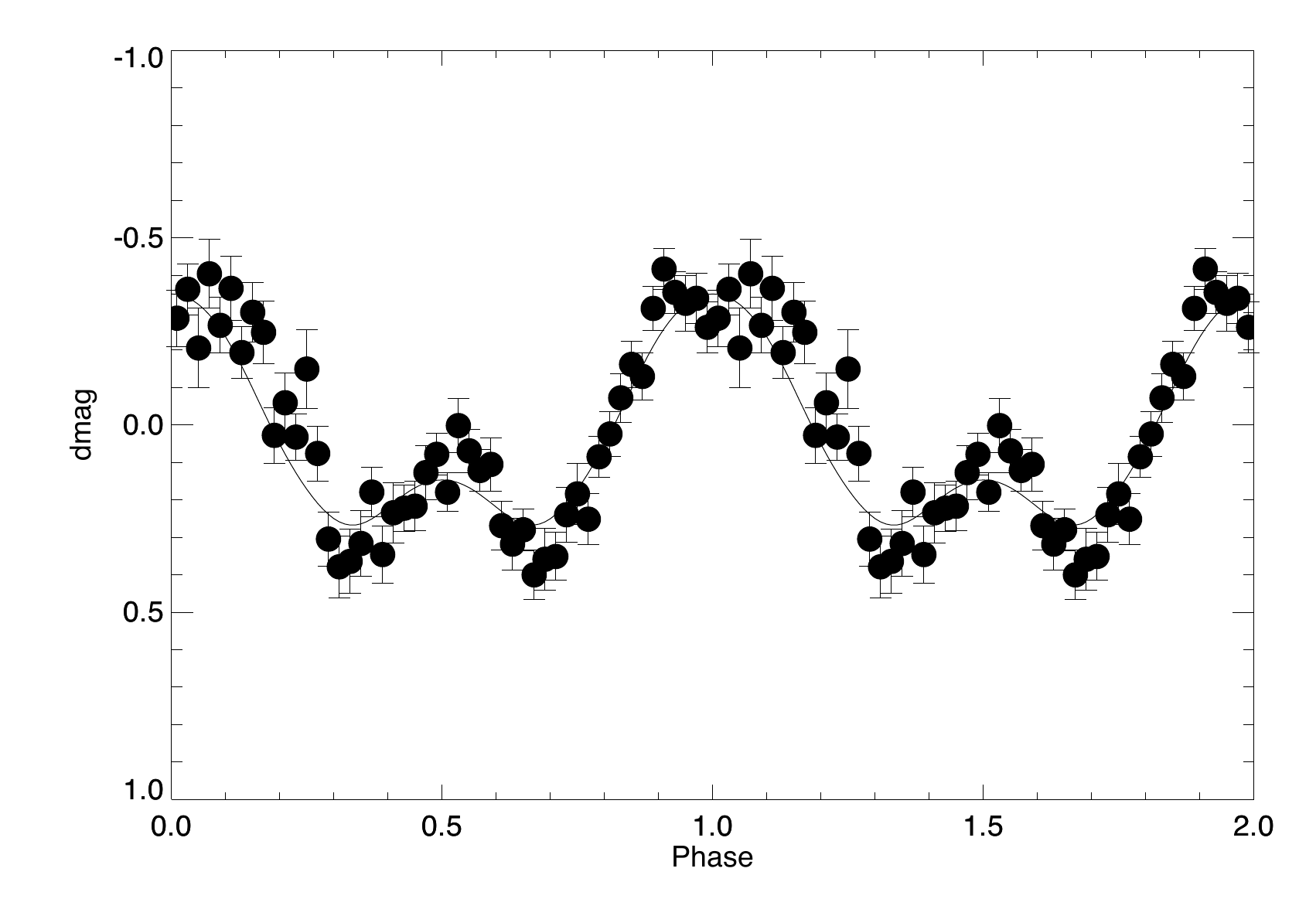}\\
    \includegraphics[width=0.8\columnwidth]{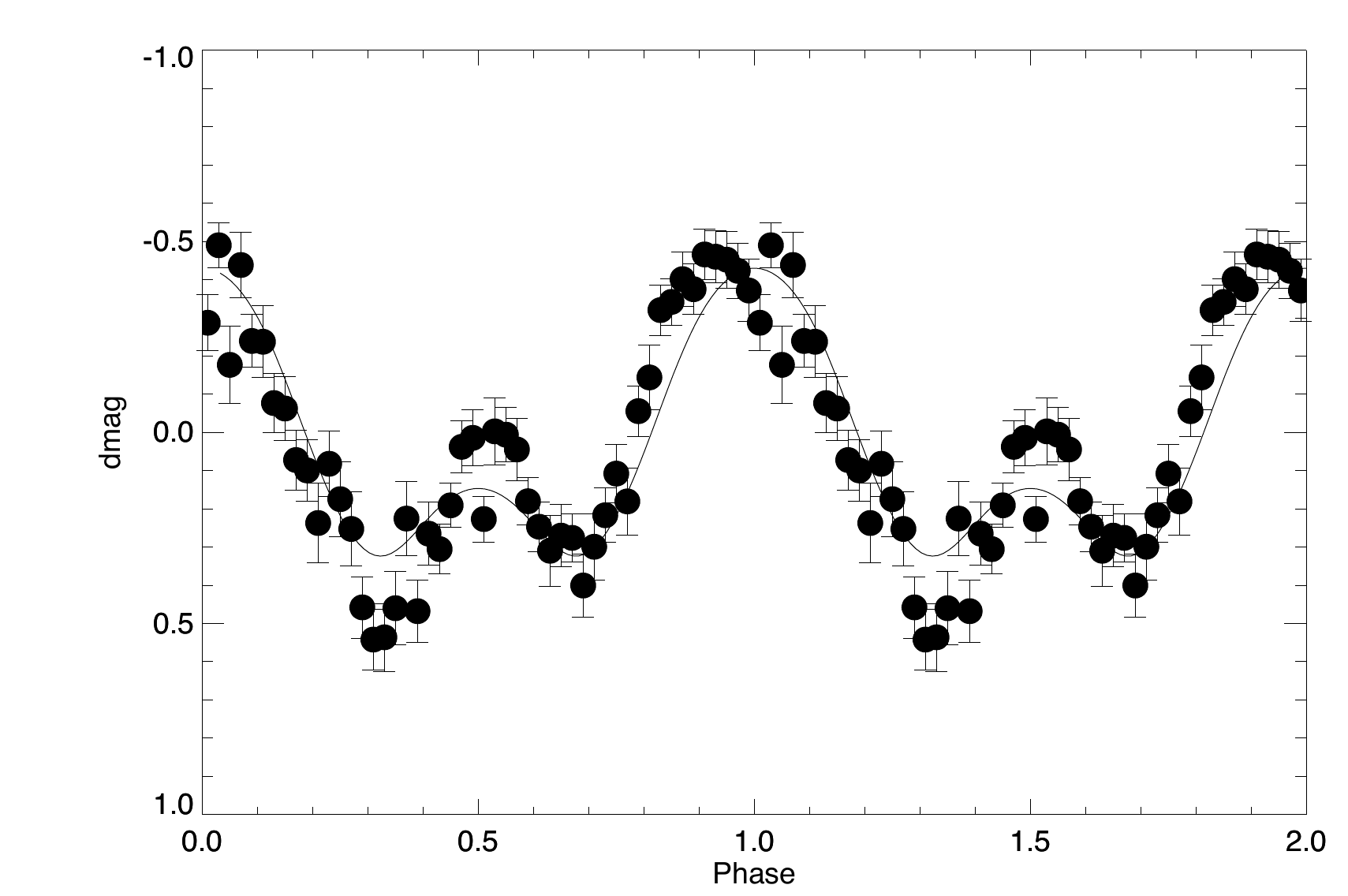}&
    \includegraphics[width=0.8\columnwidth]{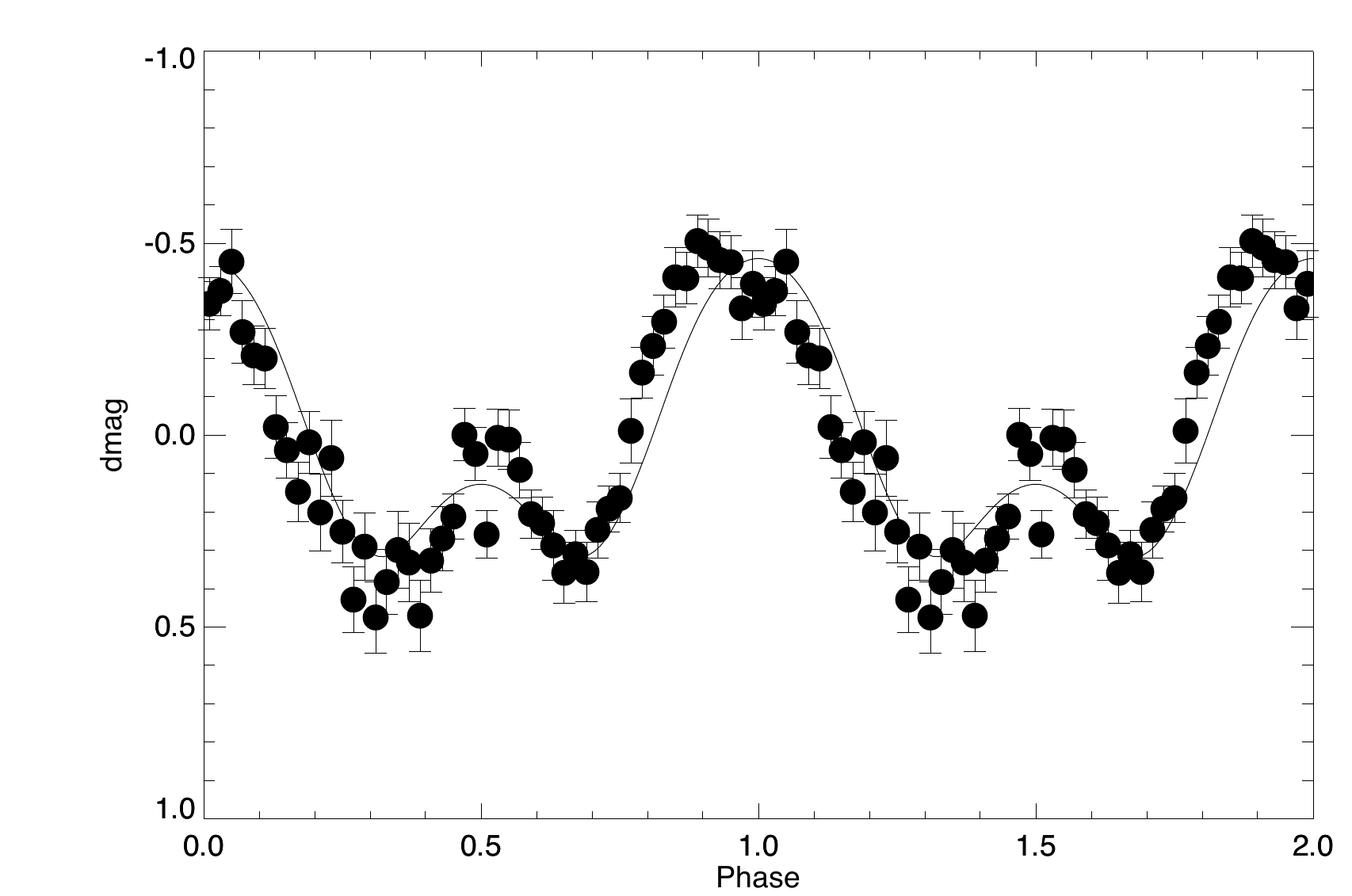}\\
    \end{tabular}
    \caption{Phased curves from various data sets. From left to right and top to bottom:
AAVSO data from 1896 to 1916; 
AAVSO data from 1906 to 1926; 
AAVSO data from 1916 to 1936; 
AAVSO data from 1926 to 1946; 
AAVSO data from 1936 to 1956; 
AAVSO data from 1946 to 1966; 
AAVSO data from 1956 to 1976; 
AAVSO data from 1966 to 1986; 
    Overimposed, the best model light curves}
    \label{fig:phased_curves_1}
\end{figure*}

\begin{figure*}[htb]
    \centering
    \begin{tabular}{cc}
    \includegraphics[width=0.8\columnwidth]{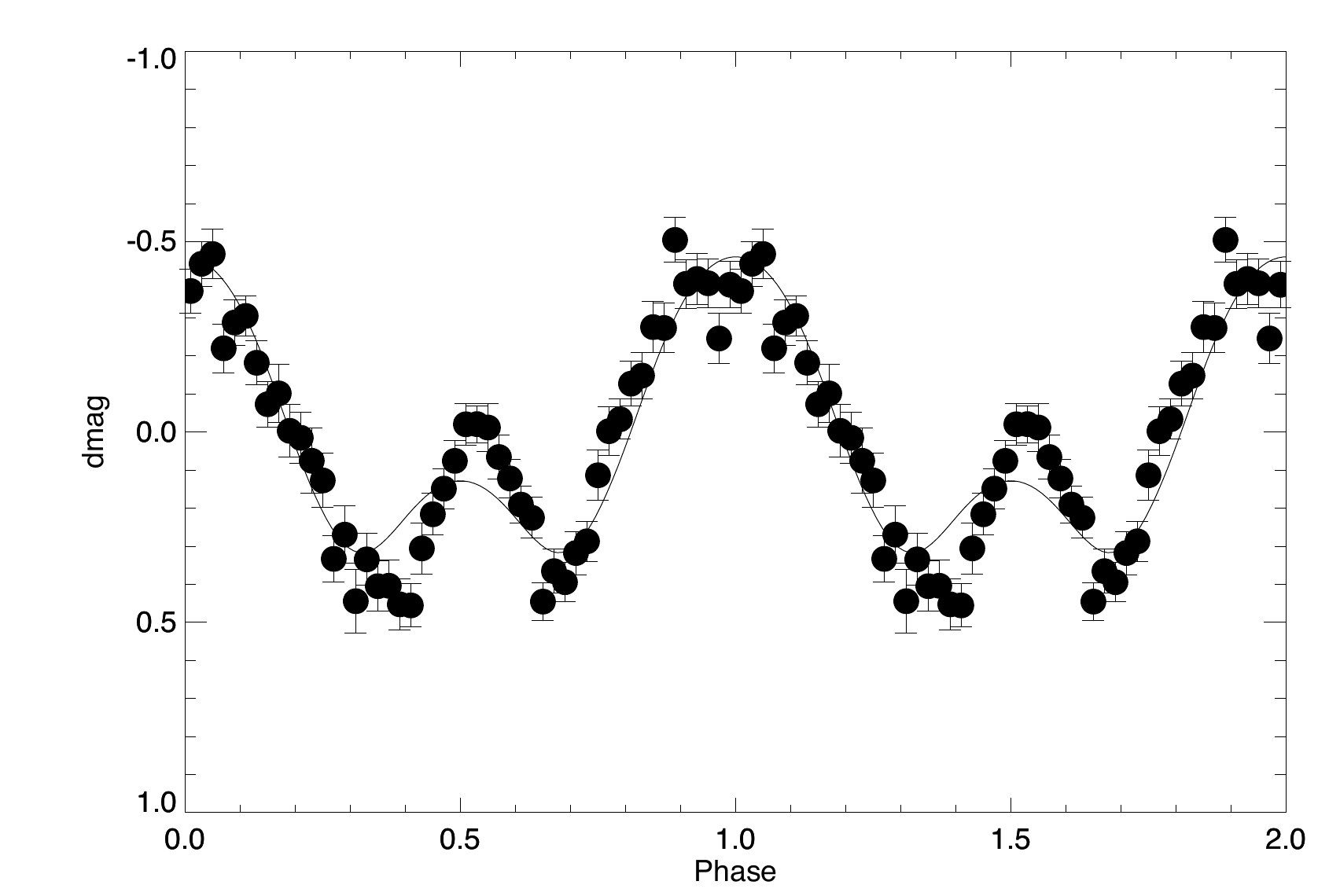}&
    \includegraphics[width=0.8\columnwidth]{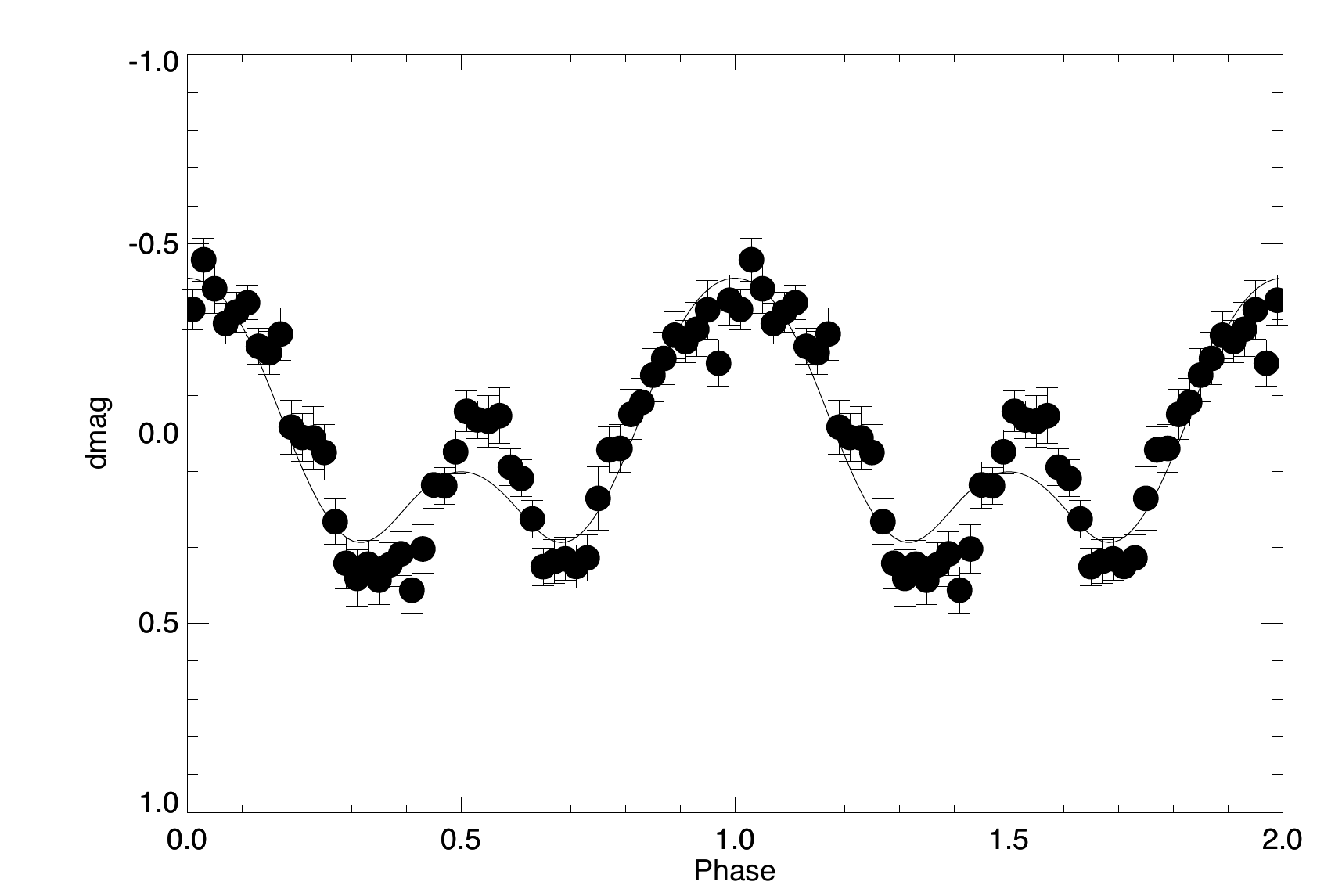}\\
    \includegraphics[width=0.8\columnwidth]{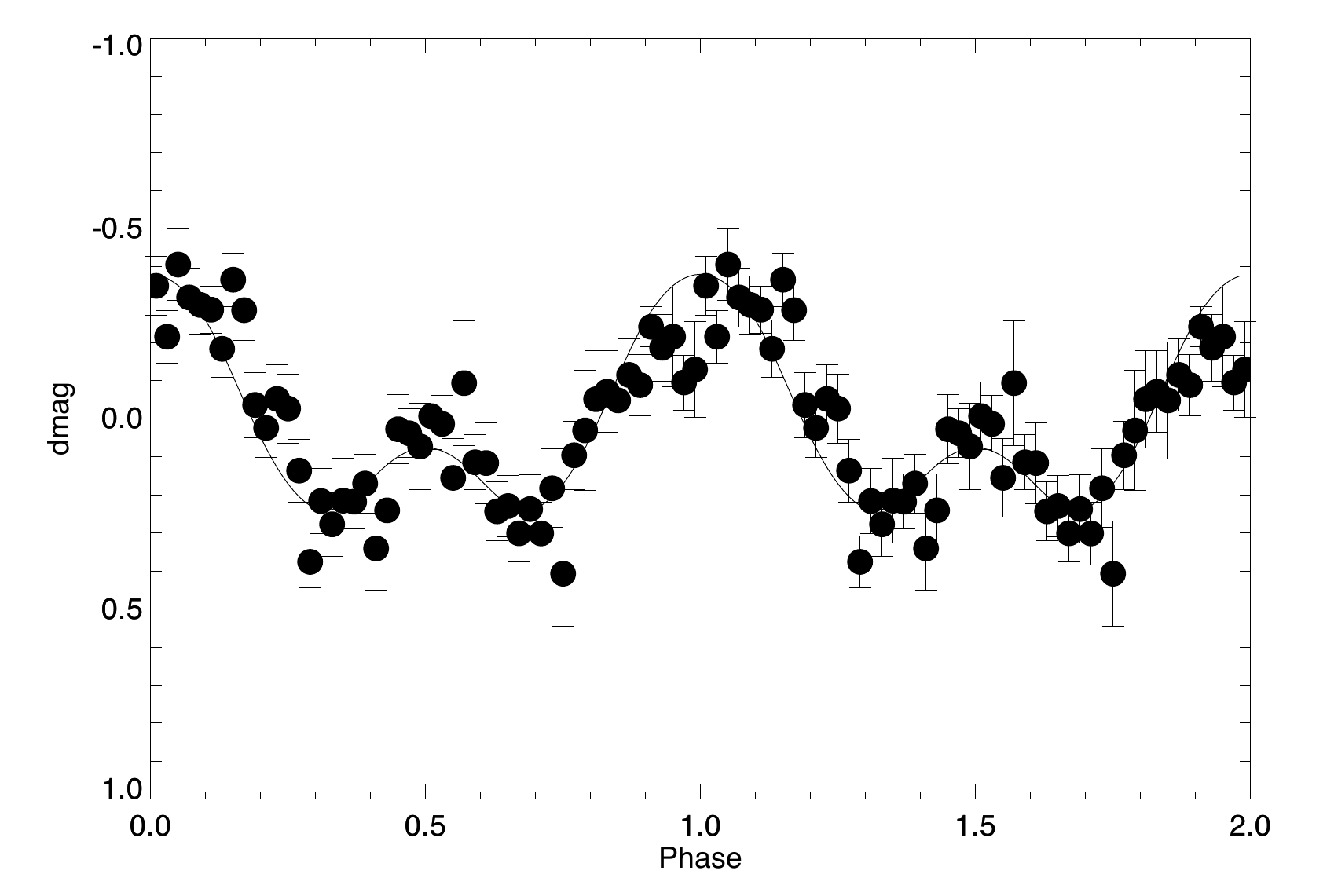}&
    \includegraphics[width=0.8\columnwidth]{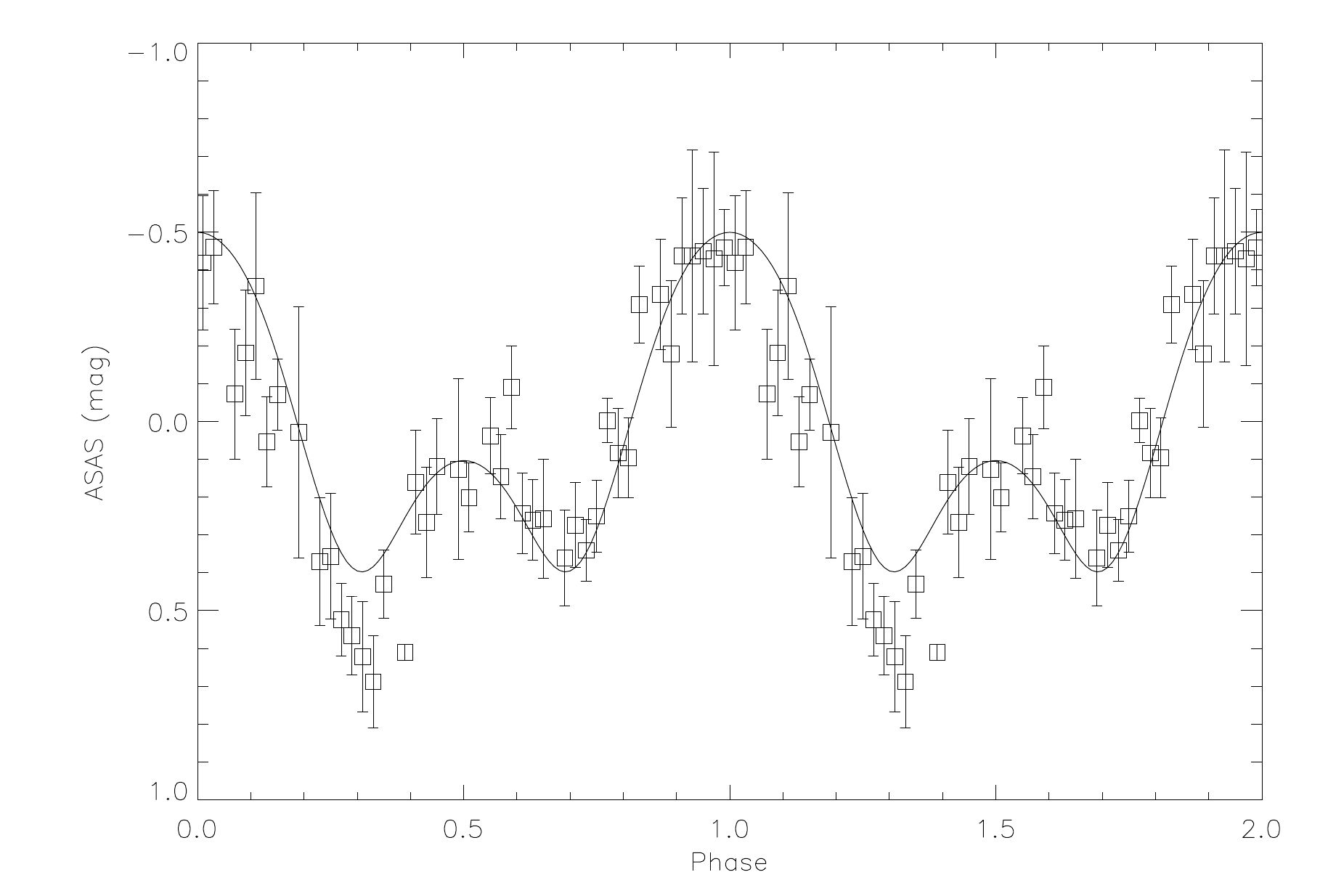}\\
    \includegraphics[width=0.8\columnwidth]{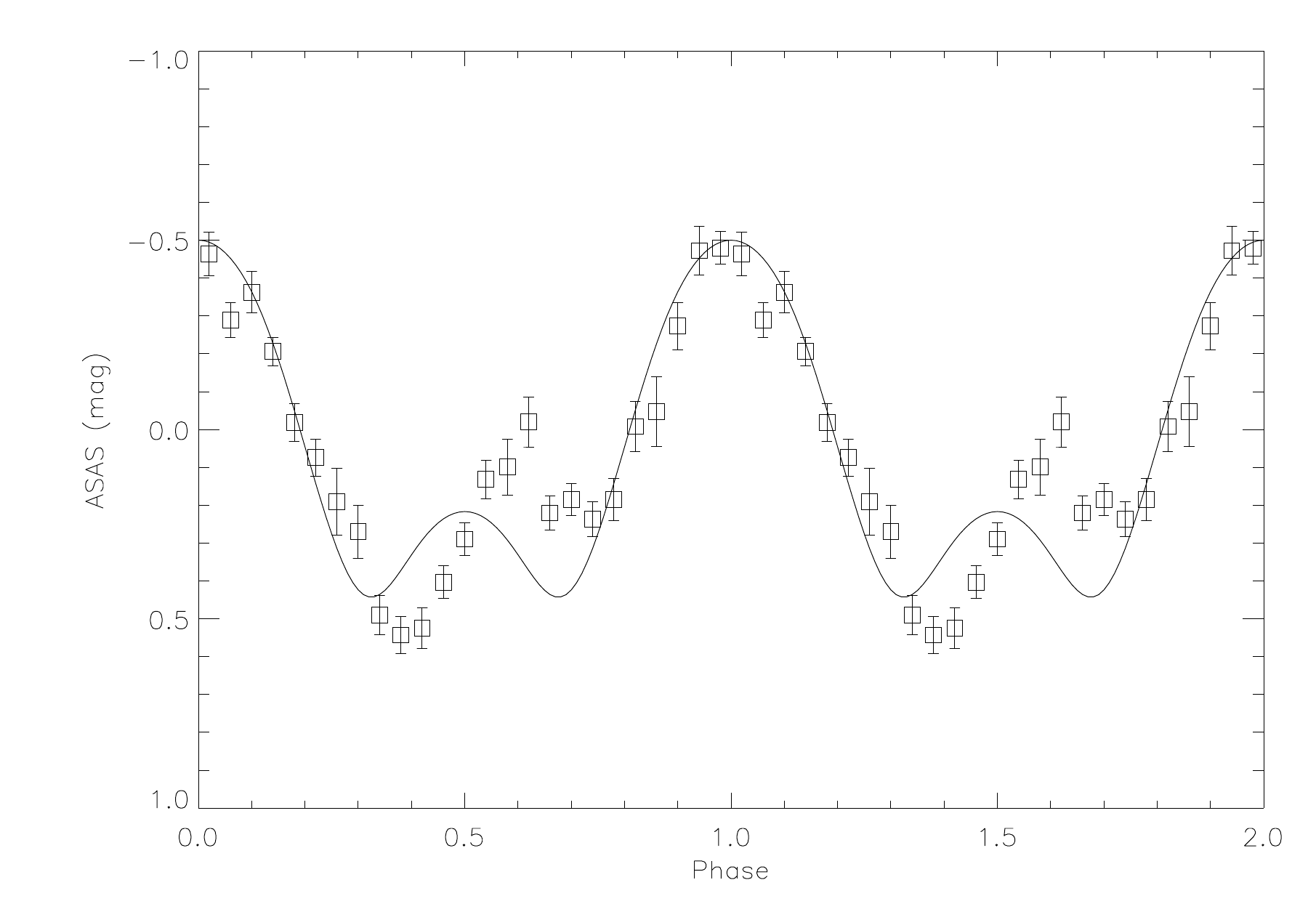}&
    \includegraphics[width=0.8\columnwidth]{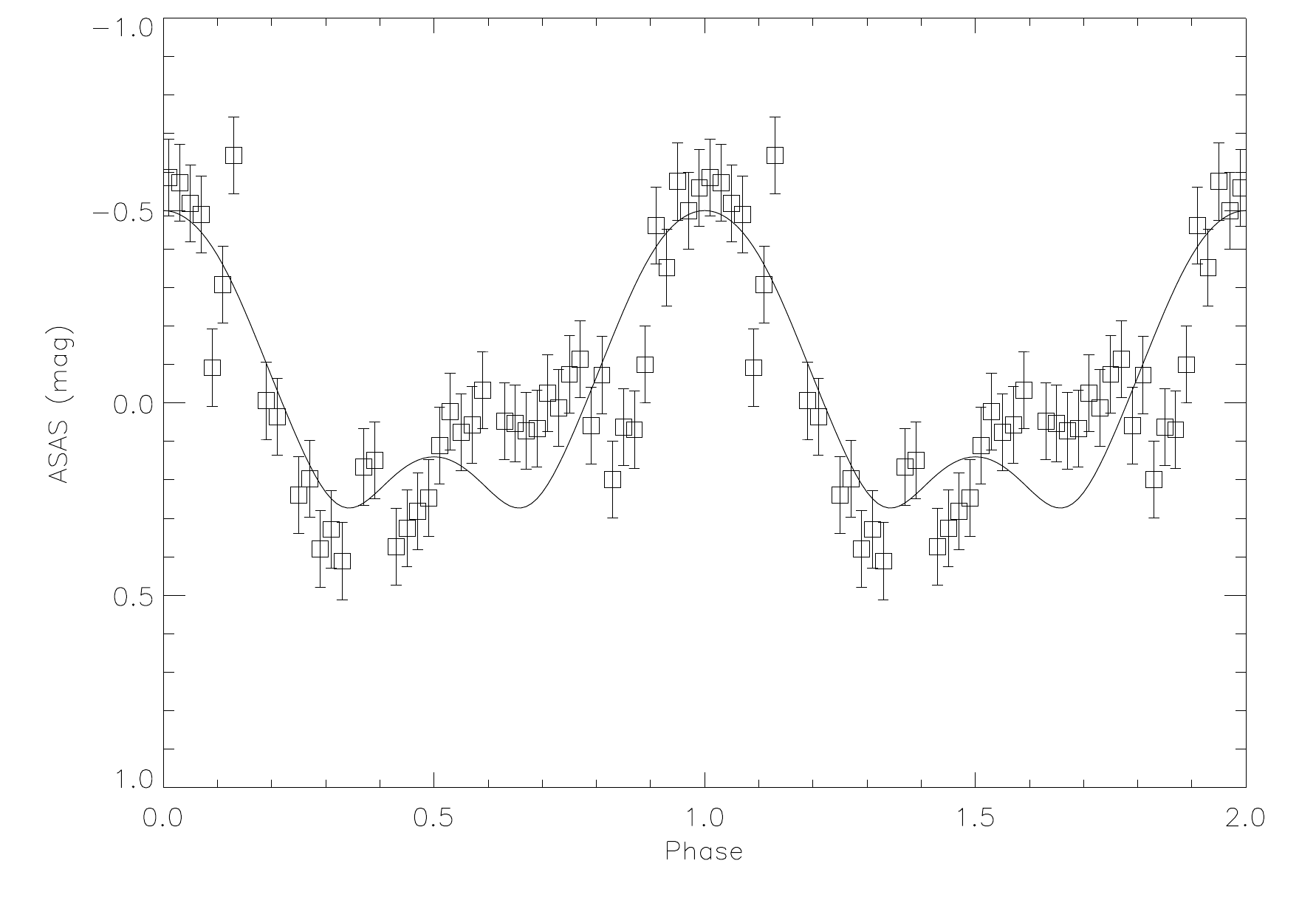}\\
    \includegraphics[width=0.8\columnwidth]{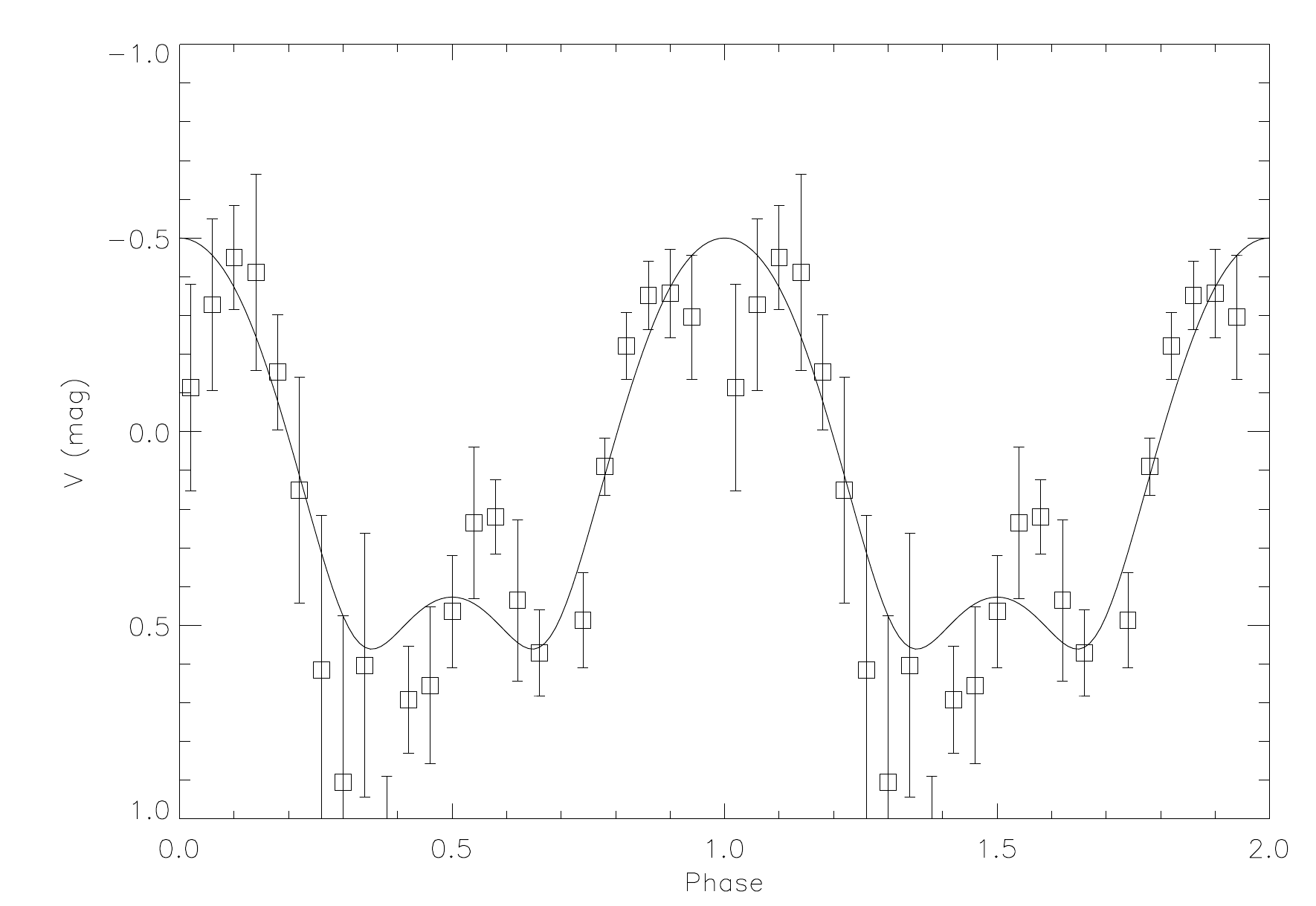}&
    \includegraphics[width=0.8\columnwidth]{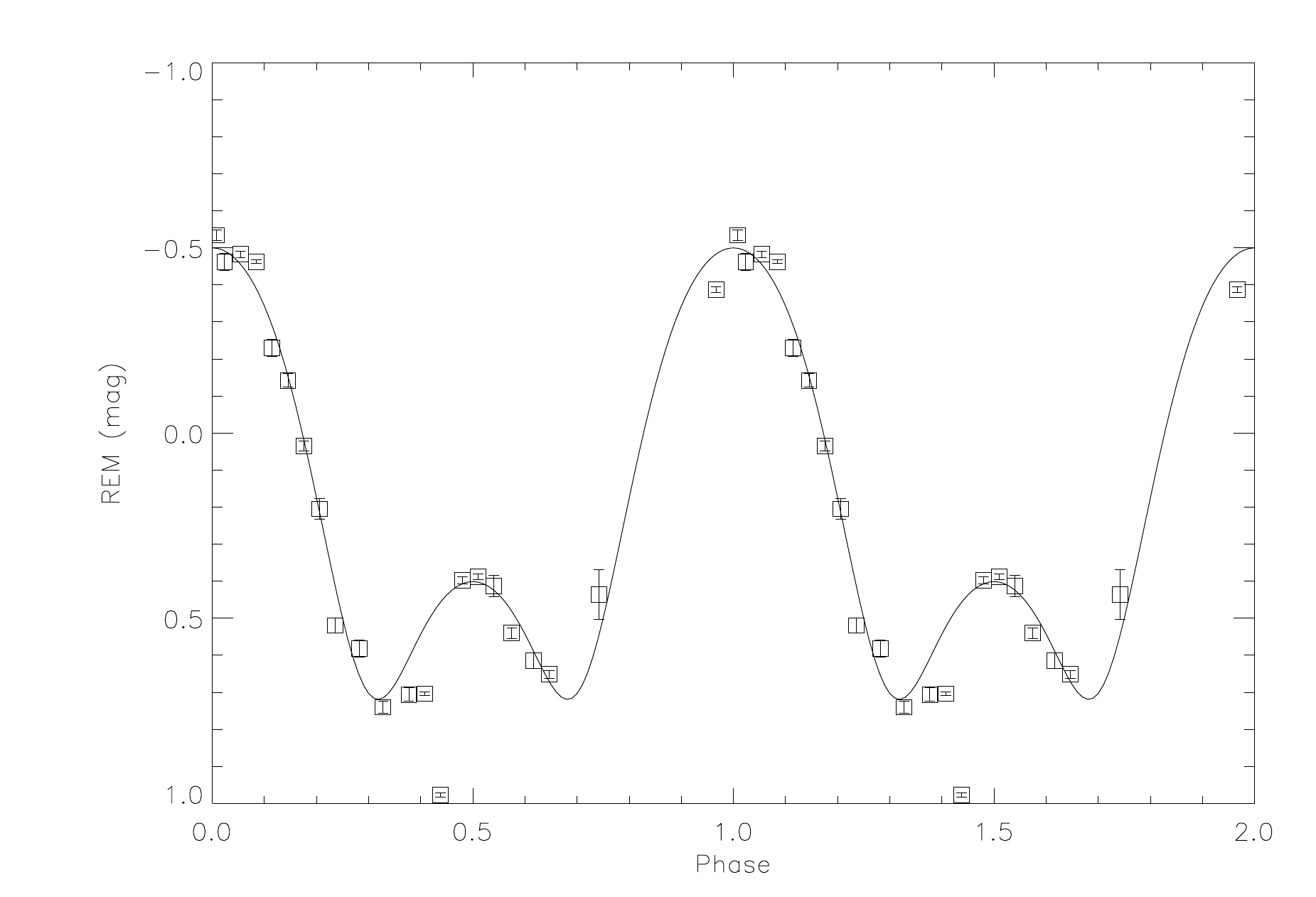}\\
    \end{tabular}
    \caption{Phased curves from various data sets. From left to right and top to bottom:
AAVSO data from 1976 to 1996; 
AAVSO data from 1986 to 2006; 
AAVSO data from 1996 to 2006; 
ASAS data;
ASAS-SN data;
SuperWasp data;
Herbst data;
REM data;
    Overimposed, the best model light curves}
    \label{fig:phased_curves-_2}
\end{figure*}

%\end{appendix}

\end{document}